\documentclass{emulateapj}

\newcommand{\teff}{T$_{\rm eff}$ }
\newcommand{\tsin}{T$_{\rm eff}$}

\begin{document}

\title{The Rise of the AGB in the Galactic Halo: 
Mg Isotopic Ratios and High Precision Elemental Abundances in M71 Giants
\altaffilmark{1} }

\author{J. Mel\'endez\altaffilmark{2} \& J.~G. Cohen\altaffilmark{3}}

\altaffiltext{1}{Based on observations obtained at the
W.M. Keck Observatory, which is operated jointly by the California
Institute of Technology, the University of California and the National
Aeronautics and Space Administration}

\altaffiltext{2}
{Centro de Astrof\'{\i}sica da Universidade do Porto, Rua das Estrelas, 4150-762 
Porto, Portugal (jorge@astro.up.pt) }

\altaffiltext{3}{Palomar Observatory, Mail Stop 105-24,
California Institute of Technology, Pasadena, California 91125
(jlc@astro.caltech.edu)}

\begin{abstract}
High-resolution (R $\approx$ 100 000), high signal-to-noise 
spectra of M71 giants have been obtained with HIRES at the Keck~I Telescope 
in order to measure their Mg isotopic ratios, as well as elemental
abundances of C, N, O, Na, Mg, Al, Si, Ca, Ti, Ni, Zr and La.
We demonstrate that M71 has
two populations, the first having weak CN, normal O, Na, Mg, and Al, and a 
low ratio of $^{26}$Mg/Mg ($\sim$4\%)  consistent with models of galactic chemical
evolution with no contribution from AGB stars.
The Galactic halo could have been formed from the dissolution
of globular clusters prior to their intermediate mass stars reaching the AGB.
The second population has
enhanced Na and Al accompanied by lower O and by higher $^{26}$Mg/Mg
($\sim$8\%), consistent with models which do incorporate ejecta
from AGB stars via normal stellar winds. 
All the M71 giants
have identical [Fe/H], [Si/Fe], [Ca/Fe], [Ti/Fe] and [Ni/Fe]
to within $\sigma = 0.04$~dex (10\%). We therefore infer
that the timescale for formation of the first generation of
stars we see today in this globular cluster must be sufficiently short
to avoid a contribution from AGB stars, i.e. less than $\sim$0.3~Gyr.
Furthermore, the Mg isotopic
ratios in the second M71 population, combined with their elemental
abundances for the light elements, demonstrate that the difference
must be the result of adding in the ejecta of intermediate mass
AGB stars.  Finally we suggest that the low amplitude of the
abundance variations of the light elements within M71 is due
to a combination of its low mass and its relatively high Fe-metallicity.
\end{abstract}


%

\section{Introduction \label{section_intro} }

Galactic globular clusters (GCs) show star-to-star variations
in the light elements including Li, C, N, O, Na, Mg, and Al discovered 
more than 30 years ago.  Correlations and anti-correlations
among O, Na, Mg, and Al are seen in essentially all GCs,
but not in field stars, as reviewed by \cite{gratton_araa}.  
These correlations are highly suggestive of proton burning
of H at high temperatures \citep{dennisenkov}.  However, detection
of these correlations among GC main sequence stars which have not
yet evolved sufficiently for their central temperatures to become
hot enough for the necessary nuclear reactions to operate by
\cite{gratton01} and by \cite{ramirez02} demonstrated definitively that these
patterns are imprinted from an external source; they cannot be the
result of internal nucleosynthesis followed by mixing to the surface
of processed material.  \cite{cohen_m15_c} have shown that superposed
on this is the imprint of mixing of material processed through H
burning via the CN cycle, but only
among the most luminous of the globular cluster giants.

The scenario generally invoked to explain this is 
that more than one generation of stars was formed in the GC; 
first one that produced the heavy elements we see today
within a system where the gas was well mixed throughout
the entire volume.  This was followed by the
second generation of stars, including low mass stars
that survive to the present, then a subsequent stellar generation
formed from gas
contaminated by the ejecta of an earlier generation
that was not uniformly mixed throughout the volume.  It is
generally presumed  that material processed within the
interiors of AGB stars, mixed to the stellar surface,
then ejected by stellar winds, is the culprit producing
the star-to-star variations detected among GC stars, see, e.g. \cite{fenner03},
\cite{cohen_m15_c}, \cite{ventura08}, and references therein.
Recently rapidly rotating  massive stars have also been invoked
for this purpose
as during their main sequence phase they transport material
processed through H burning at high temperatures to their surfaces, see, e.g.
\cite{decressin07}.  This material has an  escape velocity
low enough to be confined to the cluster,
unlike that of the subsequent radiatively driven winds or the
eventual SN, where the ejected material has such a high velocity that
it is lost by the cluster.

However, some of the details of the
behavior seen thus far do not match the predictions for 
high temperature H-burning.
The most discrepant case is that of the Mg isotopes. The first attempt
to measure Mg isotopic ratios for GC giants was by
\cite{shetrone96}, with additional recent
studies by  \cite{2003A&A...402..985Y} and by \cite{2006ApJ...638.1018Y}.
All of these authors found substantial contributions from the heavier
Mg isotopes, reaching up to half of the total Mg, far larger than the
solar ratio, which is 79:10:11 for $^{24}$Mg:$^{25}$Mg:$^{26}$Mg.
Measurement of the Mg isotopic ratios requires 
detection of weak contributions from the rarer heavier Mg isotopes
in the wings of much stronger
$^{24}$MgH lines, hence exquisite
spectra of very high spectral resolution ($\sim$100,000) and
very high signal-to-noise ratio,
as well as careful attention to details of the line
list used for the spectral synthesis and to the analysis procedure.

Here we present a new attempt to measure the Mg isotopic ratios
in the giants of the relatively metal-rich GC M71 which we hope fulfills
the criteria described above to achieve a reliable result. 
We exploit our exquisite spectra to derive
precision abundances for several additional elements in M71, including
the light elements O, Mg, Na, and Al.  We discuss our
results in the context of the history of star formation in globular
clusters, and the contribution of AGB stars to GCs and to the halo field.

\section{Observations}
The sample consists of 9 giant stars in the globular cluster M71,
eight observed with HIRES (Vogt et al. 1994) at the Keck I telescope by us 
and one (M71 A4) observed with HDS at the Subaru telescope by 
Yong et al. (2006)\footnote{The 
spectrum of M71 A4 has been kindly made available to us by D. Yong and W. Aoki}. 

The Keck observations were obtained in August 2004, June 2005 and September 2007.
All our data was obtained after the HIRES upgrade (August 2004), thus taking
advantage of improvements in efficiency, spectral coverage and spectral
resolution. A resolving power of $R \approx 10^5$ was achieved using a
0.4"-wide slit, accepting a substantial light loss at the slit in order
to achieve the necessary spectral resolution.   The signal-to-noise
level exceeded 100 per pixel, with 1.3 km s$^{-1}$/pixel.
In Fig. 1 we compare our HIRES spectrum of M71 1-77
with the HDS spectrum of M71 A4 (Yong et al. 2006), which is of similar
quality ($R \approx$ 90 000).

The spectral orders were extracted  using both IRAF and 
MAKEE\footnote{MAKEE was developed 
by T. A. Barlow specifically for reduction of Keck HIRES data.
It is freely available at
http://www2.keck.hawaii.edu/realpublic/inst/hires/data\_reduction.html}.
Further data reductions (Doppler correction, continuum normalization, and 
combining spectra) were performed with IRAF.

\section{Analysis}

Initially stellar parameters were obtained from photometry, following
the precepts set out in
\cite{cohen01}.  We are fortunate to now have
$J$ and $K_s$ photometry
from 2MASS \citep{2mass1,2mass2} (not available at the time of
the initial M71 abundance study)
and optical photometry with
more extensive coverage of
the area of M71 from \cite{2000PASP..112..925S} as updated
on his on-line database available through the CADC. 
We adopt E($B-V$) = 0.25~mag from
the Feb.~2003 version of the
online database of 
\cite{harris96}\footnote{http://physwww.mcmaster.ca/${\sim}$harris/Databases.html}.

Our preliminary analysis using the photometric stellar parameters and 
Kurucz overshooting model atmospheres (Castelli, Gratton \& Kurucz 1997)
revealed that M71 seems to show only very slight abundance variations (if any)
in some elements (e.g. Mg), and therefore very high precision is
needed to ascertain if abundance variations are present or not. 
Seeking the highest possible precision, and given the potential presence
of reddening variations across the field of M71, we decided to perform the analysis using 
spectroscopically determined stellar parameters. The stars have similar 
temperatures (\teff = 4045 $\pm$ 225 K), so this approach should 
minimize errors in the {\it relative} stellar parameters.

Excitation and ionization balances of Fe\,{\sc i} and Fe\,{\sc ii} were 
performed with the 2002 version of MOOG (Sneden 1973), 
using the Fe\,{\sc i} line list of Alves-Brito et al. (2009, in preparation)
and the Fe\,{\sc ii} line list of Hekker \& Melendez (2007),
which is described in Mel\'endez \& Barbuy (2009).
These line lists have been  carefully scrutinized to avoid significant blends
(e.g. CN) in metal-rich cool giants. All equivalent widths were
measured by hand using IRAF, employing the deblend option when necessary.
Microturbulence was obtained by flattening the trend in the iron abundance
from Fe\,{\sc i} lines versus reduced equivalent width. The zero-points of our spectroscopic 
stellar parameters have been determined by Melendez et al. (2008) using many
bright field giants with precise effective temperatures using
the infrared flux method temperature scale of Ram{\'{\i}}rez \& Mel{\'e}ndez (2005),
and with surface gravities determined with Hipparcos parallaxes and
Yonsei-Yale (Demarque et al. 2004) and Padova isochrones (da Silva et al. 2006).

Once the stellar parameters (\tsin, log $g$, $v_{mic}$) were set, the 
macroturbulence ($v_{mac}$) was determined employing five Fe\,{\sc i} lines 
(605.60, 607.85, 609.67, 612.02, 615.16~nm) and checked with other lines present 
in the same region (605 to 615~nm). Note that in previous works on Mg isotopic ratios 
(e.g. McWilliam \& Lambert 1988; Yong et al. 2003) only the Ni\,{\sc i} 511.54 nm 
and the Ti\,{\sc i} 514.55~nm were used to obtain the macroturbulent broadening.
However, we find that
these lines are not useful in metal-rich cool giants, because the region
around each of them is badly blended. The region we have chosen
to use for this purpose (605-615 nm)
is much cleaner, although we have also included in our spectrum synthesis
blends by atomic, C$_2$, and CN lines. We recommend the use of this region to
determine $v_{mac}$ in future studies of Mg isotopic ratios in relatively metal-rich very cool stars.

The line lists of MgH and C$_2$ (0-0 band) have been previously described 
in \cite{2007ApJ...659L..25M}, and the extension of our C$_2$ line list to
other bands is described in \cite{melendez_asplund}. The CN line list adopted 
here is from \cite{1999ApJS..124..527M}. For the atomic lines, we have used when
available transition probabilities based on laboratory measurements, otherwise
we have adopted astrophysical $gf$-values based on Arcturus.  We only used
the best, most isolated atomic lines within the wavelength range covered
by our spectra, listed in Table~\ref{table_linelist}.

The Mg isotopic ratios were determined by spectral synthesis, as described
in \cite{2007ApJ...659L..25M}, except that in cool metal-rich giants the
MgH feature at 513.46 nm (usually employed for Mg isotopic ratios; e.g. Gay \& Lambert 2000)
seems blended, so instead we use the 513.43 nm feature. Two other MgH features
at 513.87 and 514.02 nm (e.g. Yong et al. 2003; Melendez \& Cohen 2007) were
also used for obtaining the Mg isotopic ratios. After the first trials, it was clear that 
the $^{24}$Mg:$^{25}$Mg:$^{26}$Mg ratios were no larger than terrestrial (79:10:11), 
therefore we have computed synthetic spectra with isotopic ratios ranging
from $^{24}$Mg:$^{25}$Mg:$^{26}$Mg = 100:0:0 to 76:12:12.

The isotopic ratios were determined by $\chi^2$ fits, by computing
$\chi^2$ = $\Sigma$($O_i - S_i$)/$\sigma^2$, where $O_i$ and $S_i$ represents
the observed and synthetic spectrum, respectively, and $\sigma$ = (S/N)$^{-1}$.
Examples of the $\chi^2$ fits are shown in Figs. 2 and 3 for the three recommended
MgH features. The resulting Mg isotope ratios are presented in Tab.~2.
In this table are also given the standard deviation from the three different
MgH features, which for $^{25}$Mg and $^{26}$Mg have a typical value 
of $\sigma$ = 1.7\% and 1.9\%, respectively, corresponding to
standard errors of 1.0\% and 1.1\%. As discussed below, the
true error for $^{26}$Mg is actually somewhat smaller than for $^{25}$Mg.
Due to the larger isotopic separation of the $^{26}$MgH lines,
the determination of $^{26}$Mg/$^{24}$Mg is more reliable than $^{25}$Mg/$^{24}$Mg.

The elemental abundances of C, N, O, Na, Mg and Al were obtained by spectral
synthesis, taking into account blends by atomic, C$_2$ and CN lines,
and the abundances of Si, Ca, Ti, Ni and Zr were obtained from equivalent widths.
Carbon abundances were obtained from C$_2$ lines around 563 nm
and checked using CH lines around 430 nm (Plez et al. 2008, see also Plez \& Cohen 2005). 
Nitrogen abundances were obtained using CN lines around 630-637 nm and 669-671 nm. 
The CN-rich stars have N abundances $\sim$3 times higher than the CN-weak
stars, but the difference may be higher because TiO blends prevent a precise 
determination of N abundances in the CN-weak stars.
The elemental abundance ratios are given in Tabs.~2 and 3.

The iron abundances, [C,N,O/Fe] abundance ratios, and 
the C+N and C+N+O abundance sums, are shown in Fig. 4 as a function of effective temperature.
A similar plot for the isotopic $^{24,25,26}$Mg/Mg, $^{25,26}$Mg/$^{24}$Mg ratios 
and the elemental abundance ratios [Na,Mg,Al/Fe] is presented in Fig. 5,
and the [Si,Ca,Ti,Ni,Zr,La/Fe] ratios are shown in Fig. 6.

In these and subsequent figures
CN-weak and CN-rich giants are represented by open and filled circles,
respectively. The CN status of our sample stars have been obtained from 
Smith \& Norris (1982) and Lee (2005), except for star M71~I, for which no previous
information on its CN bands is available in the literature. Since this star shows 
strong CN bands in our HIRES spectrum, we have also included it as a CN-strong star.

As discussed below, some abundance ratios show trends with \tsin.
The coefficients of the fits for the CN-weak stars are given in Tab.~4.

\section{Discussion}

\subsection{Correlations between $^{24,25,26}$Mg/Mg, O, Na, Mg and Al}

The Mg isotopic ratios and the elemental abundances of O, Na, Mg and Al
for the CN-weak stars seem to show some trend with effective temperature,
probably due to NLTE and 3D effects (Asplund 2005; Collet et al. 2007).
After correcting for trends with \tsin, the scatter in [O,Na,Mg,Al/Fe] for the
CN-weak stars is only 0.018, 0.039, 0.018 and 0.021 dex, respectively.
The CN-rich stars depart from the behavior of the CN-weak stars, 
showing larger $^{25,26}$Mg/Mg ratios, larger Na and Al elemental abundances,
and lower O and Mg elemental abundances.

In Figs. 7-11 are shown the $^{24}$Mg/Mg, $^{25}$Mg/Mg, 
$^{25}$Mg/$^{24}$Mg, $^{26}$Mg/Mg and $^{26}$Mg/$^{24}$Mg ratios 
as a function of [O,Na,Mg,Al/Fe]. In these plots the trend with effective temperature
(Figs. 4-5, Tab. 4) has been corrected in the [O,Na,Mg,Al/Fe] ratios, to prevent spurious trends. 
The $^{25,26}$Mg/$^{24}$Mg isotopic ratios show a larger spread 
than $^{25,26}$Mg/Mg due to the spread in $^{24}$Mg/Mg 
(CN-rich giants show lower $^{24}$Mg/Mg than CN-weak giants).

In Fig. 7 we show that $^{24}$Mg in CN-rich stars is anticorrelated with
Na and probably also with Al.
The $^{25}$Mg/Mg ratios are less reliable than the $^{26}$Mg/Mg 
ratios. Although Table 2 suggest higher uncertainties for $^{26}$Mg 
than for $^{25}$Mg, with standard errors of 1.1\% and 1.0\%, respectively,
the star-to-star scatter for the CN-weak stars actually shows
that the errors for $^{25}$Mg and $^{26}$Mg are 0.9\% and 0.8\%,
respectively, i.e., in both cases somewhat lower than the errors
estimated from Table 2. The higher error bar ($\pm$0.9\%) for $^{25}$Mg 
is expected due to the smaller isotopic shift of the $^{25}$MgH lines.
Hence, $^{25}$Mg does not show clear trends
besides the fact that $^{25}$Mg is enhanced in CN-rich giants (Figs. 8-9).
The CN-weak stars do not show any correlation between $^{26}$Mg and 
the elemental abundances of Na, O, Mg and Al (Figs. 10-11), 
but the CN-rich giants show strong, weak and 
no correlations between $^{26}$Mg and the elemental 
abundances of Na and Al, O, and Mg, respectively.

Fig.~12 shows the correlations between the elemental abundances
of O, Na, Mg and Al, which have been corrected for trends with \teff (Figs. 4-5, Tab.~4),
and therefore any remaining trend or scatter should be real. 
Interestingly, even though the CN-weak stars show only
small scatter in their abundances ratios (0.02--0.04 dex), and seem to be
roughly constant between the uncertainties, there is a hint of
a correlation between Na and Al. On the other hand, Al and Mg, and O and Na,
seem to be anti-correlated, although the evidence is weaker for the O:Na
anti-correlation. The CN-strong stars show strong correlation and anti-correlation
between Na and Al, and O and Na, respectively.

\subsection{C+N and C+N+O in M71 giants}
As can be seen in Fig. 4, the C+N abundance sum is larger in CN-rich 
than in  CN-weak M71 giants. This is mainly due to the large N enhancement
of the CN-rich giants, which have N abundances $\sim$0.5 dex higher
than the CN-weak giants. Keck low resolution spectra of CN-weak and CN-strong
main sequence stars in M71 show that CN-strong dwarfs are also enhanced 
in nitrogen when compared to CN-weak dwarfs (Briley \& Cohen 2001).

The C+N+O abundance sum is constant within 0.1 dex (Fig. 4), 
in agreement with other high resolution analysis of
GC giants in the literature, which find C+N+O constant within 0.3 dex for 
NGC 6712 ([Fe/H] = -1.0, Yong et al. 2008), 
M4 ([Fe/H] = -1.1, Smith et al. 2005) and 
M13 ([Fe/H] = -1.5, Cohen \& Mel\'endez 2005). 

Recently, Yong et al. (2009) have found a large C+N+O spread of
0.57 dex for NGC 1851 giants ([Fe/H] = -1.2). This spread exceeds 
their estimated uncertainty (0.14 dex), and is much larger than that
observed in other GCs, where no significant spread is found. 
Yong et al. (2009) interpret the large spread in C+N+O 
as a signature of AGB pollution in NGC 1851, 
and they conclude that if the AGB scenario is applicable 
to other GCs with constant C+N+O, then perhaps the masses of the
AGB polluting stars in NGC 1851 were lower than in other GCs.
This is probably also the reason why NGC 1851, unlike other GCs, 
shows a large spread in the s-process elements Zr and La (Yong et al. 2009).

\subsection{Chemical evolution of the Mg isotopes}

As discussed in \cite{2007ApJ...659L..25M}, the lightest and most abundant
Mg isotope can be formed in massive stars \citep[e.g.][]{woosley95}  as
a primary isotope from H.  The heavier isotopes are formed as secondary
isotopes, as well as in intermediate-mass AGB stars \citep{karakas08},
so the isotopic ratios $^{25,26}$Mg/$^{24}$Mg increase with the
onset of AGB stars.  Therefore Mg isotopic ratios in halo stars can be
used as a chronometer
to constrain the rise of the AGB in our Galaxy.  In our earlier
paper we applied this technique to a sample of metal-poor Galactic
halo field dwarfs to establish that the AGB did not make
a significant contribution to Mg in the Galactic halo 
at least until [Fe/H] $\sim -1.3$~dex was reached.

Fig.~13 shows the chemical evolution of  $^{26}$Mg/$^{24}$Mg for
halo field dwarfs (Melendez \& Cohen 2007), adding in the M71 giants,
with models 
from \cite{fenner03} for the solar neighborhood
both with and without the contribution of
AGB stars.  We see that the M71 CN-weak giants have very low
Mg isotopic ratios $^{26}$Mg/Mg ($\sim$4\%)  consistent with models of 
galactic chemical
evolution with no contribution from AGB stars, and extending our previous
results based on metal-poor halo field dwarfs to still higher
metallicity.  Based on the calculations of stellar yields for AGB stars 
by \cite{karakas03}, updated in \cite{karakas08}, stars of initial mass 
3--6 M$_{\odot}$ are
required to contribute significant amounts of the heavier Mg isotopes.
Stars in this mass range reach
the upper AGB at an age of $\sim$0.3~Gyr; see, e.g., 
the Padova evolutionary tracks\footnote{http://pleiadi.pd.astro.it/}.
For a GC as metal-rich as is
M71, this timescale must reflect
the minimum age difference between the CN-weak and the CN-rich
stars in the GC as well as the maximum range in age of the CN-weak
stars we see today in M71, assuming uniform mixing of the gas
within the GC. Furthermore, the enrichment of the halo gas
from which M71 eventually formed must have
reached such a high Fe-metallicity without
a substantial contribution from proton burning of H at high
temperatures.  We argue that AGB stars are the source
of the ``polluted'' material, rather than it coming from
rapidly rotating massive stars, on the
basis of timescales, which would be uncomfortably short
in the latter case, and also the uniformity of the
heavy elements within all stars of M71.   Our precision abundances
define an upper limit to the range of abundances of the heavy
elements in M71;  all the M71 giants
have identical [Fe/H], [Si/Fe], [Ca/Fe], [Ti/Fe] and [Ni/Fe]
to within $\sigma = 0.04$~dex (10\%) after the trends with \teff\
are removed (see Figs.~4 and 6).  This would be hard to achieve
if massive stars were the source of polluting gas in GCs.

On the other hand, the
M71 giants with strong CN (accompanied by lower O with higher Na and Al 
than those of the weak CN giants) show higher
Mg isotopic ratios $^{26}$Mg/Mg ($\sim$8\%), but these are still
quite low compared
to previous studies by \cite{2006ApJ...638.1018Y}
mostly in GCs of lower metallicity.
In particular, star M71~A4 has been also analyzed by
\cite{2006ApJ...638.1018Y},
who find higher isotopic ratios than ours.
This must be due to the different analysis techniques, as the
same observed spectrum was used. In particular, note that
their macroturbulent velocity for this star is lower.
As shown in Fig.~13, our newly measured Mg isotope ratios for the CN-rich
giants in M71 lie on the predicted
curve for models which include the appropriate contribution to the chemical
inventory from material processed through high temperature H-burning
which we believe comes from AGB stars during the normal course
of evolution of intermediate mass stars.

\subsection{Neutron capture processes}

Fig. 6 displays [La/Fe] vs \teff\ for the M71 sample
of luminous giants.
The 6390~\AA\ La~II line was synthesized with 
hyperfine structure and transition probability
from \cite{lawler01}.  The results are more uncertain
than for most of the other elements discussed here
as there is only one La~II line which was sufficiently
unblended to be used, and it lies within a CN band.
However, there appears to be a slight excess of La
among the CN-strong M71 giants. This excess is small
enough that it would not have been detected in previous
analyses such as that of \cite{ramirez02}, whose spectra
were of lower spectral resolution and SNR, with consequent
larger uncertainties.

La, together with Ba\footnote{The Ba~II lines
are too strong to be able to detect the small 
star-to-star variations seen here}, are the archetypical elements
indicating a contribution to the star's chemical inventory via
$s$-process neutron capture. But this provides no discrimination 
in the source
of the $s$-process, be it intermediate mass AGB stars or
young massive rotating stars.  The former, especially
in metal-poor AGB stars, is well known as a site
for such production \citep{busso01}, while very recent
work by the NuGrid project \citep{hirschi08} demonstrates
that the $s$-process can operate efficiently in rotating young
massive stars.

Zr is also believed to be produced primarily in AGB stars 
via the $s$-process of neutron capture
on Fe seeds.  However, we do not see any detectable difference
for the mean [Zr/Fe] between the CN-rich and the CN-weak M71 giants,
while differences in the isotopic ratio $^{26}$Mg/Mg are apparent,
and we believe there are differences in [La/Fe] as well.
There are four detected Zr~I lines, so the lack of detectable
star-to-star variation in [Zr/Fe]
must be regarded as well established .
The production of Zr in AGB stars is discussed by \cite{busso01}
and more recently by \cite{travaglio04}.  The yield of Zr via the main $s$-process
depends strongly
on the size of the $^{13}$C pocket, as this isotope is the source
of the neutrons for this channel.
The $^{13}$C is itself produced  locally within the AGB star
by proton captures on the abundant isotope $^{12}$C, but the
requirement of having substantial C together with protons  in
a region hot enough for these reactions to proceed leads
to these reactions occurring under very constrained conditions,
as described in detail in \cite{busso01}.
The weak $s$-process, also involved in Zr production, utilizes
$^{22}$Ne as the neutron source.
\cite{karakas08} provides an estimate of the total yield
of elements heavier than the Fe-peak (i.e. beyond the maximum atomic mass
included in their reaction
network) in AGB stars.  Their results 
suggest that the production
of Zr is biased towards lower mass AGB stars than those within which the
heavy isotopes of Mg are made, which could explain why [Zr/Fe] does not
appear to vary among the M71 giants in our sample.  Better
models to verify the consistency of our results with the 
predicted behavior of
Zr and La vs the Mg isotope ratios in  AGB stars as a function of 
mass are needed.

Eu serve the same role for the $r$-process of neutron capture which
produces selected isotopes of heavy elements beyond the Fe peak.
No credible star-to-star variations could be detected among
the luminous M71 giants
in [Eu/Fe] based on the strength of 
the 6645~\AA\ line of Eu~II.

\section{Conclusions}

With the aid of exquisite spectral resolution and very high signal-to-noise
spectra combined with a very careful analysis, we have determined
Mg isotopic ratios for 9 luminous giants in the metal-rich Galactic
globular cluster M71.  We have also used these spectra to determine
precision abundances for several other elements in this GC, including
the light elements O, Na, Mg, and Al. 

The abundances of Si, Ca, Ti, Ni and Fe do not show any star-to-star variations.
The total range for the absolute Fe abundance, log[$\epsilon$(Fe)], 
among the sample of 9 giants
in M71 is only 0.08~dex. Once the dependence on \teff\ is
removed, all the M71 giants
have identical [Fe/H], [Si/Fe], [Ca/Fe], [Ti/Fe] and [Ni/Fe]
to within $\sigma = 0.04$~dex (10\%). This places a strong constraint
on the uniformity of mixing in the young GC.
These elements cannot
have been produced in anything other than the first generation
of GC stars.

We see the expected correlations and anticorrelations among the light elements
O, Na, Mg, and Al in the M71 giants.
But the amplitude of the star-to-star variations among these elements is small,
0.3~dex for [O/Fe], 0.6~dex for [Na/Fe], 0.2~dex for [Al/Fe], and
at most 0.1~dex for [Mg/Fe].  \cite{carretta07} have looked for
a link between chemical anomalies along the RGB and other properties
of GCs.  In addition to the obvious suggestion that higher amplitude
star-to-star variations should be found  in higher mass GCs,
which  with their higher binding energies may be better able to retain 
stellar ejecta, they suggest that the high temperature extension of
the horizontal branch blue tail is longer in GCs with higher amplitude
Na-O anticorrelations.  The latter is attributed to a spread in He,
and hence may again be tied to the ability to retain ejected gas,
see the discussion in \cite{carretta07}.  We suggest that one
additional parameter is relevant, the Fe-metallicity of the GC.
As the stellar metallicity becomes higher, the effect of the addition
of AGB processed material, whose nucleosynthesis proceeds among
the light elements in a manner which is somewhat independent of
their initial metallicity, would be diluted, and the resulting yield,
assumed to be positive, reduced. Such an effect is apparent
in the calculation of AGB yields by \cite{karakas08}, among others.
This would explain
why the most prominent cases of star-to-star variations within GCs
are seen among the lower metallicity GCs (i.e. M15, M13, etc) 
even though there are a number
of quite massive high-metallicity GCs such as 47~Tuc 
for which \cite{carretta04} found only modest star-to-star variations
among the light elements.
In this context it is interesting to note that the behavior of
CH and CN bands in these relatively high metallicity GCs tends
to be bimodal, while in more metal-poor GCs, stars fill the entire
range of C and N abundances without being so concentrated
towards the upper and lower extremes 
of [C/Fe] \citep[see, e.g. Fig.~11 of][]{cohen_m15_c}.

The heavy Mg isotopes among the  CN weak giants in our M71 sample have very low
Mg isotopic ratios $^{26}$Mg/Mg ($\sim$4\%) which are  consistent with models of 
galactic chemical
evolution with no contribution from AGB stars.  These stars
are both CN weak and have light element abundances typical of field
Galactic halo metal-poor stars of similar Fe-metallicity;
we call them the 
``normal'' stars.  \cite{kroupa02} discuss forming the Galactic halo field
via dissolved GCs; our results suggest that this must have involved
the  ``normal'' GC stars from
clusters which dissolved early on before  their intermediate-mass stars
reached the AGB.
The CN strong
M71 giants have higher Mg isotopic ratios $^{26}$Mg/Mg ($\sim$8\%),
but these are far below previously published values by
\cite{2006ApJ...638.1018Y} and references therein for more metal-poor GC giants.
This group of stars, in addition to a higher fraction of the heavier Mg isotopes,
shows enhanced Na and Al, accompanied by lower O abundances.

The behavior of the Mg isotopes in the ``normal'' stars is
reproduced  by models of galactic chemical evolution by \cite{fenner03}
without any contribution from AGB stars.  Their models with
the AGB contribution expected  in the normal course of stellar evolution
reproduce the behavior of the heavy Mg isotope rich M71 giants.
These stars must represent a later generation of stars
formed sufficiently long after the first that the AGB contribution
to their chemical inventory was included. 
Alternatively, the CN-rich and CN-weak stars may have formed at
the same time, but the CN-weak stars could have been formed in 
an unmixed environment, while the CN-strong stars were born from 
material polluted by AGB stars.

With the present work, we believe we have demonstrated convincingly
that the difference between the generations of GC stars is consistent 
in detail with the
contribution of AGB stars.  Furthermore, we have
extended our previous
results from \cite{2007ApJ...659L..25M} 
based on metal-poor halo field dwarfs to still higher
metallicity, up to the Fe-metallicity of M71.  
Based on the calculations of stellar yields for AGB stars 
by \cite{karakas03}, updated in \cite{karakas08}, stars of initial mass 
3--6 M$_{\odot}$ are
required to contribute significant amounts of the heavier Mg isotopes.
Stellar evolutionary tracks establish that  
stars in this mass range reach the upper AGB at an age of $\sim$0.3~Gyr.
Assuming uniform mixing within the gas in this GC,
for a GC as metal-rich as is M71, this timescale must reflect
the minimum age difference between the CN-weak and the CN-rich
stars in the GC and the maximum age range of the CN-weak
stars we see today. Furthermore, the enrichment of the halo gas
from which M71 eventually formed must have
reached such a high Fe-metallicity without
a substantial contribution 
of  ``polluted'' material, whose source we argue on the
basis of the tight constraints we have placed on any variation in abundance
among the Fe-peak elements in M71 must be AGB stars.

\begin{acknowledgements}

The entire Keck/HIRES user community owes a huge debt to
Jerry Nelson, Gerry Smith, Steve Vogt, and many other
people who have worked to make the
Keck Telescope and HIRES a reality and to operate and
maintain the Keck Observatory. We are grateful to the
W. M.  Keck Foundation for the vision to fund
the construction of the W.~M.~Keck Observatory.  The authors wish 
to extend
special thanks to those of Hawaiian ancestry on whose sacred mountain
we are privileged to be guests.  Without their generous hospitality,
none of the observations presented herein would
have been possible.

The authors are grateful to NSF grant AST-0507219 and 
FCT (project PTDC/CTE-AST/65971/2006 and Ciencia 2007) for partial support.
We thank B. Plez for providing the CH line list,
and D. Yong and W. Aoki for sharing their spectrum of M71 A4.
This publication makes use of data from the Two Micron All-Sky Survey,
which is a joint project of the University of Massachusetts and the 
Infrared Processing and Analysis Center, funded by the 
National Aeronautics and Space Administration and the
National Science Foundation.

\end{acknowledgements}

\clearpage

\begin{deluxetable}{ccccl}
\tablewidth{0pc}
\tablecaption{The List of Atomic Lines Used \label{table_linelist} }
\tablehead{
\colhead{Wavelength} &
\colhead{Exc. Pot.} & \colhead{log($gf$)} & \colhead{Species} \\
\colhead{(\AA)} & \colhead{(eV)} & \colhead{(dex)}  }
\startdata
 6300.30 & 0.000 & $-$9.72  &[\ion{O}{1}]  \\
 6363.77 & 0.020 &$-$10.19  &[\ion{O}{1}]  \\
 6154.22 & 2.102 & $-$1.547 &\ion{Na}{1}  \\
 6160.74 & 2.104 & $-$1.246 &\ion{Na}{1}  \\
 6318.71 & 5.108 & $-$1.945 &\ion{Mg}{1}  \\
 6319.23 & 5.108 & $-$2.165 &\ion{Mg}{1}  \\
 6696.01 & 3.143 & $-$1.481 &\ion{Al}{1}  \\
 6698.66 & 3.143 & $-$1.782 &\ion{Al}{1}  \\
 5488.98 & 5.614 & $-$1.69  &\ion{Si}{1}  \\
 6142.48 & 5.619 & $-$1.41  &\ion{Si}{1}  \\
 5590.11 & 2.521 & $-$0.571 &\ion{Ca}{1}  \\
 5867.56 & 2.933 & $-$1.57  &\ion{Ca}{1}  \\
 6156.02 & 2.521 & $-$2.42  &\ion{Ca}{1}  \\
 6166.43 & 2.521 & $-$1.142 &\ion{Ca}{1}  \\
 5453.64 & 1.443 & $-$1.61  &\ion{Ti}{1}  \\
 5648.56 & 2.495 & $-$0.25  &\ion{Ti}{1}  \\
 5913.71 & 0.021 & $-$4.10  &\ion{Ti}{1}  \\
 5918.53 & 1.066 & $-$1.47  &\ion{Ti}{1}  \\
 6092.79 & 1.887 & $-$1.378 &\ion{Ti}{1}  \\
 6273.38 & 0.021 & $-$4.17  &\ion{Ti}{1}  \\
 6706.29 & 1.502 & $-$2.78  &\ion{Ti}{1}  \\
 5435.85 & 1.986 & $-$2.60  &\ion{Ni}{1}  \\
 5468.10 & 3.847 & $-$1.61  &\ion{Ni}{1}  \\
 5589.35 & 3.898 & $-$1.14  &\ion{Ni}{1}  \\
 5846.99 & 1.676 & $-$3.21  &\ion{Ni}{1}  \\
 6086.28 & 4.266 & $-$0.51  &\ion{Ni}{1}  \\
 6108.11 & 1.676 & $-$2.44  &\ion{Ni}{1}  \\
 6130.13 & 4.266 & $-$0.96  &\ion{Ni}{1}  \\
 6127.45 & 0.154 & $-$1.06  &\ion{Zr}{1}  \\
 6134.55 & 0.000 & $-$1.28  &\ion{Zr}{1}  \\
 6140.53 & 0.519 & $-$1.41  &\ion{Zr}{1}  \\
 6143.20 & 0.071 & $-$1.10  &\ion{Zr}{1}  \\
 6390.48 & 0.321 & $-$1.41  &\ion{La}{2} \\

\enddata
\end{deluxetable}

\clearpage

\begin{table*}
\begin{minipage}[t]{\textwidth}
\caption{Stellar parameters, Mg isotopic ratios, C, N, O, C+N, and C+N+O abundances. 
The scatter ($\sigma$) of the $^{25,26}$Mg/Mg ratios obtained between the three MgH features 
is given between parenthesis.
The first four stars are CN-strong giants and the other five stars are CN-weak giants.}
\label{parameters}
\centering
\begin{tabular}{lllrrrrrrrrrrr} 
\hline\hline                
Star  & \tsin,log $g$,[Fe/H],$v_{mic}$,$v_{mac}$,$^{25}$Mg/Mg,$^{26}$Mg/Mg & [C/Fe] & [N/Fe] & [O/Fe] & A(C) & A(N) & A(O) & A(CN) & A(CNO)\\
\hline    
{}        &  (K, dex, dex, km s$^{-1}$, km s$^{-1}$, \%, \%) & (dex) & (dex) & (dex) & (dex) & (dex) & (dex) & (dex) & (dex) \\
\hline    
1-45 &  3870, 0.60, $-$0.83, 1.65, 6.0, 8.4(1.5), 9.2(2.3)  & 0.16 &   0.75 & 0.32 & 7.76 & 7.75 & 8.26 & 8.06 & 8.47 \\
I    &  4080, 1.20, $-$0.78, 1.52, 5.2, 6.8(1.7), 6.9(1.9)  & 0.13 &   0.60 & 0.49 & 7.78 & 7.65 & 8.48 & 8.02 & 8.61 \\
1-66 &  4150, 1.50, $-$0.79, 1.49, 5.6, 6.9(1.9), 7.7(2.5)  & 0.12 &   0.86 & 0.45 & 7.76 & 7.90 & 8.43 & 8.14 & 8.61 \\
1-53 &  4150, 1.50, $-$0.78, 1.52, 5.6, 6.9(0.8), 8.1(1.6)  & 0.01 &   0.91 & 0.40 & 7.66 & 7.96 & 8.39 & 8.14 & 8.58 \\
1-46 &  3820, 0.45, $-$0.81, 1.65, 6.0, 5.9(1.7), 6.0(2.2)  & 0.18 &$-$0.03 & 0.39 & 7.81 & 7.00 & 8.36 & 7.87 & 8.48 \\
A4   &  3870, 0.50, $-$0.81, 1.52, 5.5, 3.9(1.7), 5.0(2.1)  & 0.17 &   0.28 & 0.42 & 7.79 & 7.30 & 8.38 & 7.91 & 8.51 \\
1-77 &  3900, 0.55, $-$0.80, 1.57, 5.9, 5.2(1.0), 5.7(1.2)  & 0.17 &   0.22 & 0.40 & 7.80 & 7.25 & 8.37 & 7.91 & 8.50 \\
1-64 &  4160, 1.40, $-$0.76, 1.54, 5.2, 5.9(2.1), 6.0(1.8)  & 0.05 &   0.38 & 0.49 & 7.72 & 7.45 & 8.50 & 7.91 & 8.60 \\
1-21 &  4270, 1.45, $-$0.84, 1.65, 6.0, 5.2(1.9), 3.9(1.6)  & 0.19 &   0.39 & 0.58 & 7.78 & 7.38 & 8.51 & 7.93 & 8.61 \\
\hline                                 
\end{tabular}
\end{minipage}
\end{table*}

\begin{table*}
\begin{minipage}[t]{\textwidth}
\caption{[Na,Mg,Al,Si,Ca,Ti,Ni,Zr,La/Fe] abundance ratios.
The first four stars are CN-strong giants and the other five stars are CN-weak giants.}
\label{namgalsicatinizrla}
\centering
\begin{tabular}{lrrrrrrrrrrrrr} 
\hline\hline                
Star  & [Na/Fe] & [Mg/Fe] & [Al/Fe] & [Si/Fe]  & [Ca/Fe] & [Ti/Fe] & [Ni/Fe] & [Zr/Fe] & [La/Fe]\\
\hline    
{}     & (dex) & (dex) & (dex) & (dex) & (dex) & (dex) & (dex) & (dex) & (dex)\\
\hline    
1-45 &   0.48 & 0.21 & 0.36 & 0.20 & 0.18 & 0.29 & $-$0.02 &   0.23 & 0.37 \\
I    &$-$0.03 & 0.15 & 0.20 & 0.19 & 0.16 & 0.18 & $-$0.04 &   0.06 & 0.29 \\
1-66 &   0.17 & 0.14 & 0.25 & 0.26 & 0.16 & 0.17 & $-$0.01 &   0.08 & 0.36?\\
1-53 &   0.26 & 0.17 & 0.28 & 0.22 & 0.19 & 0.22 &    0.00 &   0.12 & 0.43?\\
1-46 &   0.02 & 0.25 & 0.26 & 0.16 & 0.10 & 0.27 & $-$0.06 &   0.16 & 0.30 \\
A4   &   0.00 & 0.24 & 0.26 & 0.15 & 0.18 & 0.29 & $-$0.02 &   0.17 & 0.29 \\
1-77 &   0.08 & 0.19 & 0.30 & 0.12 & 0.22 & 0.33 & $-$0.02 &   0.23 & 0.28 \\
1-64 &   0.01 & 0.18 & 0.23 & 0.18 & 0.20 & 0.22 & $-$0.04 &   0.12 & 0.29 \\
1-21 &$-$0.09 & 0.18 & 0.18 & 0.24 & 0.15 & 0.12 & $-$0.04 &$-$0.05 & 0.23 \\
\hline                                 
\end{tabular}
\end{minipage}
\end{table*}

\begin{table*}
\begin{minipage}[t]{\textwidth}
\caption{Abundance trends with temperature (A = C + slope * \tsin) for the CN-weak giants}
\label{trends}
\centering
\begin{tabular}{lrrc} 
\hline\hline                
Abundance   & C & slope \\
\hline    
{}     & (dex) & (dex K$^{-1}$)  \\
\hline    
{[Fe/H]}   & $-$0.7243 & $-$1.9401E-05  \\
{[C/Fe]}   &    0.5388 & $-$9.6626E-05  \\
{[N/Fe]}   & $-$2.4973 &    6.8564E-04  \\
{[O/Fe]}   & $-$1.0946 &    3.8726E-04  \\
{[Na/Fe]}  &    0.8767 & $-$2.1798E-04  \\
{[Mg/Fe]}  &    0.7654 & $-$1.3923E-04  \\
{[Al/Fe]}  &    1.0106 & $-$1.9097E-04  \\
{[Si/Fe]}  & $-$0.5890 &    1.8957E-04   \\
{[Ca/Fe]}  &    0.0252 &    3.6139E-05  \\
{[Ti/Fe]}  &    1.6833 & $-$3.5899E-04  \\
{[Ni/Fe]}  & $-$0.0085 & $-$6.8475E-06  \\
{[Zr/Fe]}  &    1.9213 & $-$4.4838E-04   \\
{[La/Fe]}  &    0.7269 & $-$1.1193E-04  \\
{A(C+N)}   &    7.6377 &    6.6801E-05  \\
{A(C+N+O)} &    7.3416 &    2.9914E-04  \\
\hline                                 
\end{tabular}
\end{minipage}
\end{table*}

\begin{figure}
\includegraphics[width=\hsize]{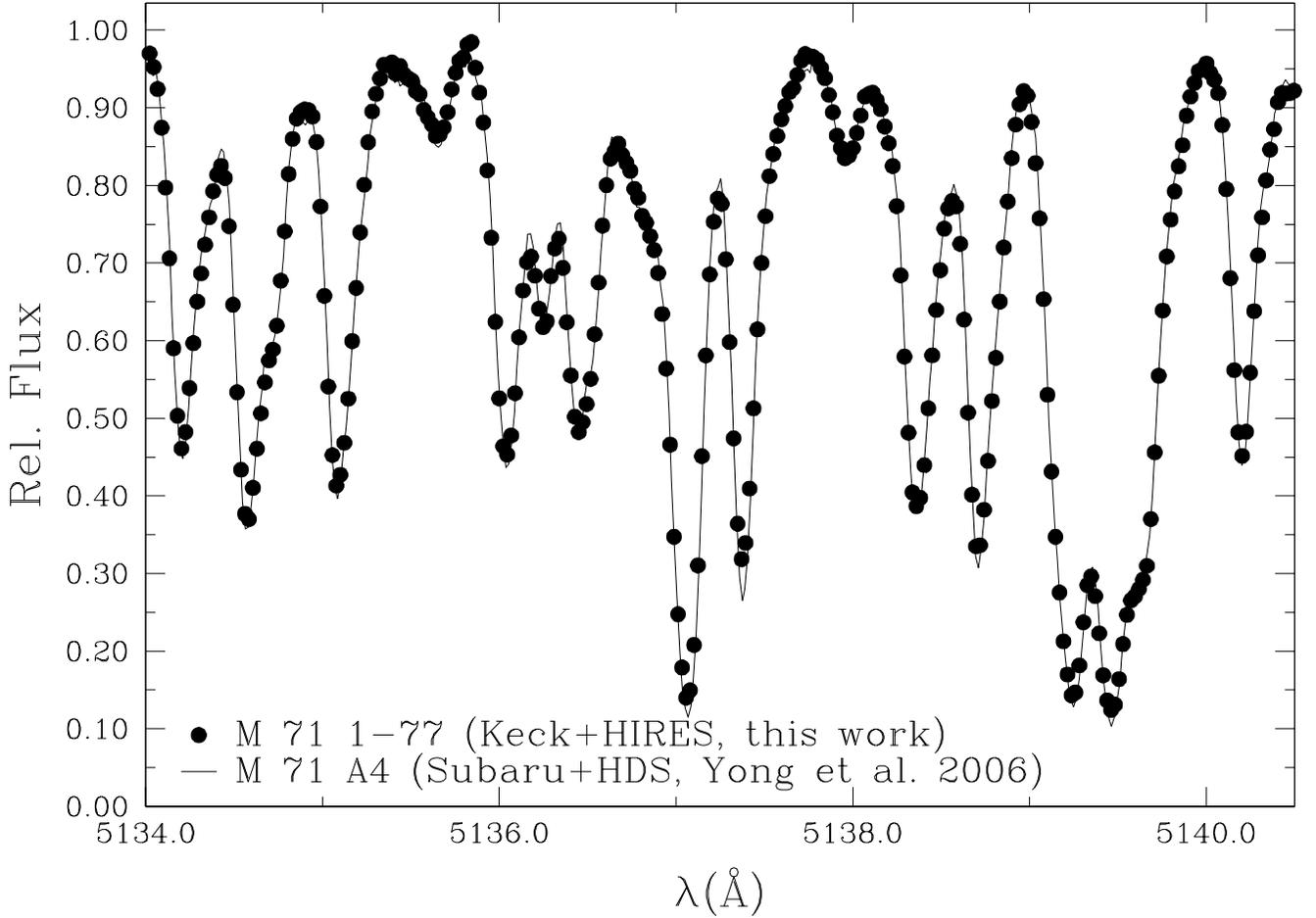}
\caption{Comparison of M71 1-77 (Keck I + HIRES, this work) and M71 A4 (Subaru + HDS, 
Yong et al. 2006). Both stars have similar stellar parameters (\teff $\approx$ 3900 K),
except that A4 has a lower macroturbulence. Both spectra were taken with a resolving
power of $\approx 10^5$.
}                      
\label{comp_A4_177}    
 \end{figure}

\begin{figure}
\includegraphics[width=5.5cm]{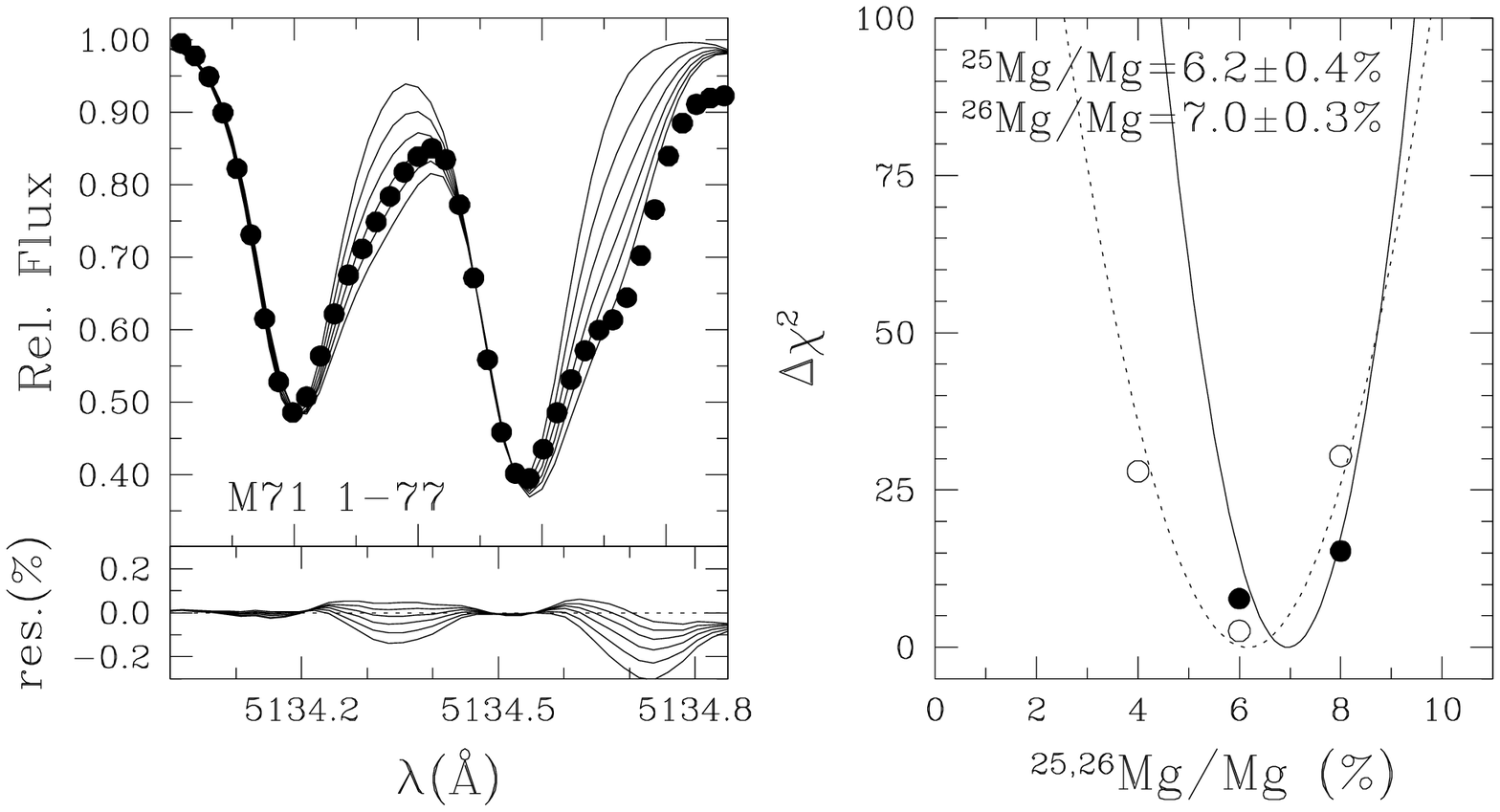}
\includegraphics[width=5.5cm]{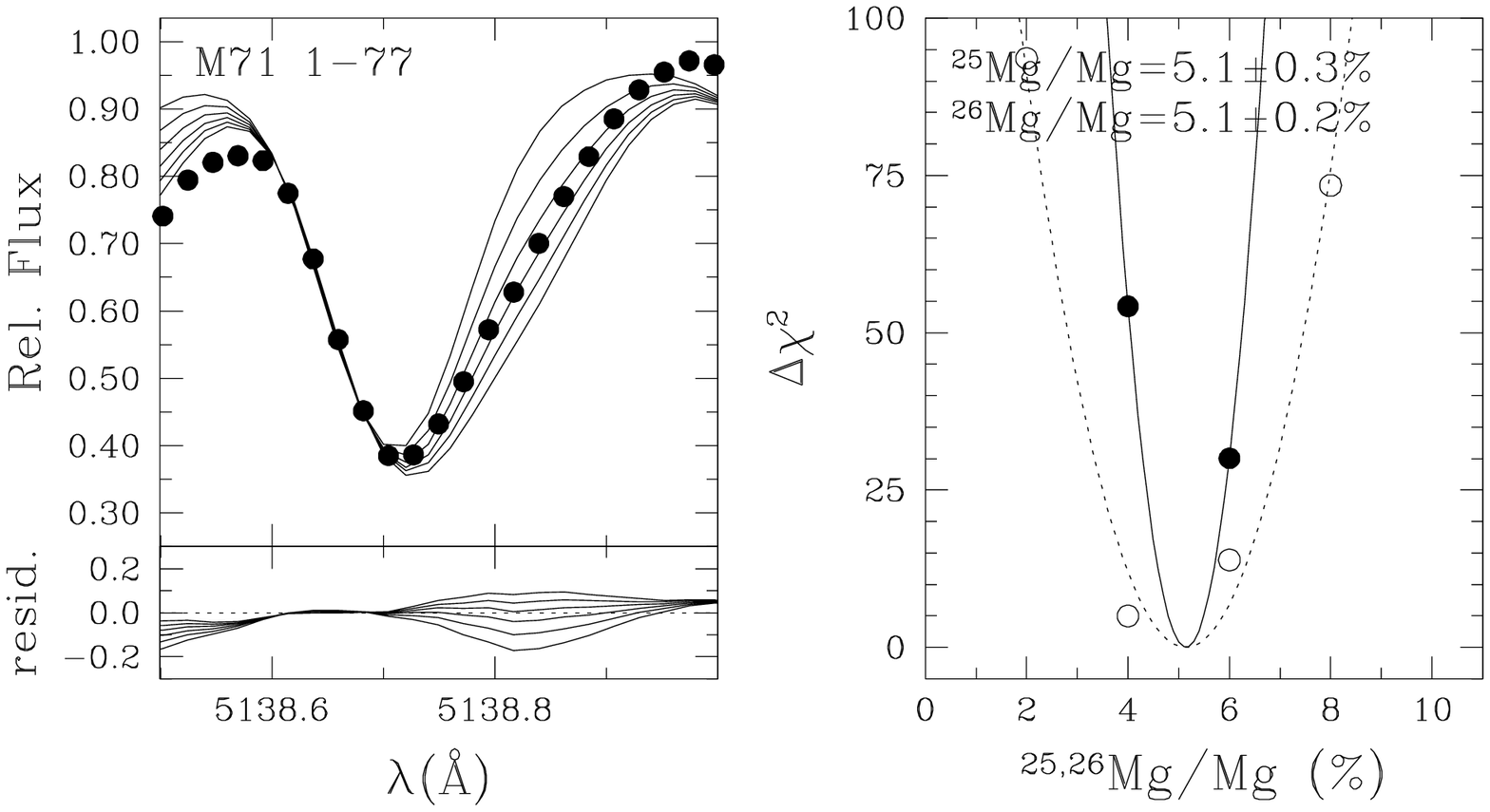}
\includegraphics[width=5.5cm]{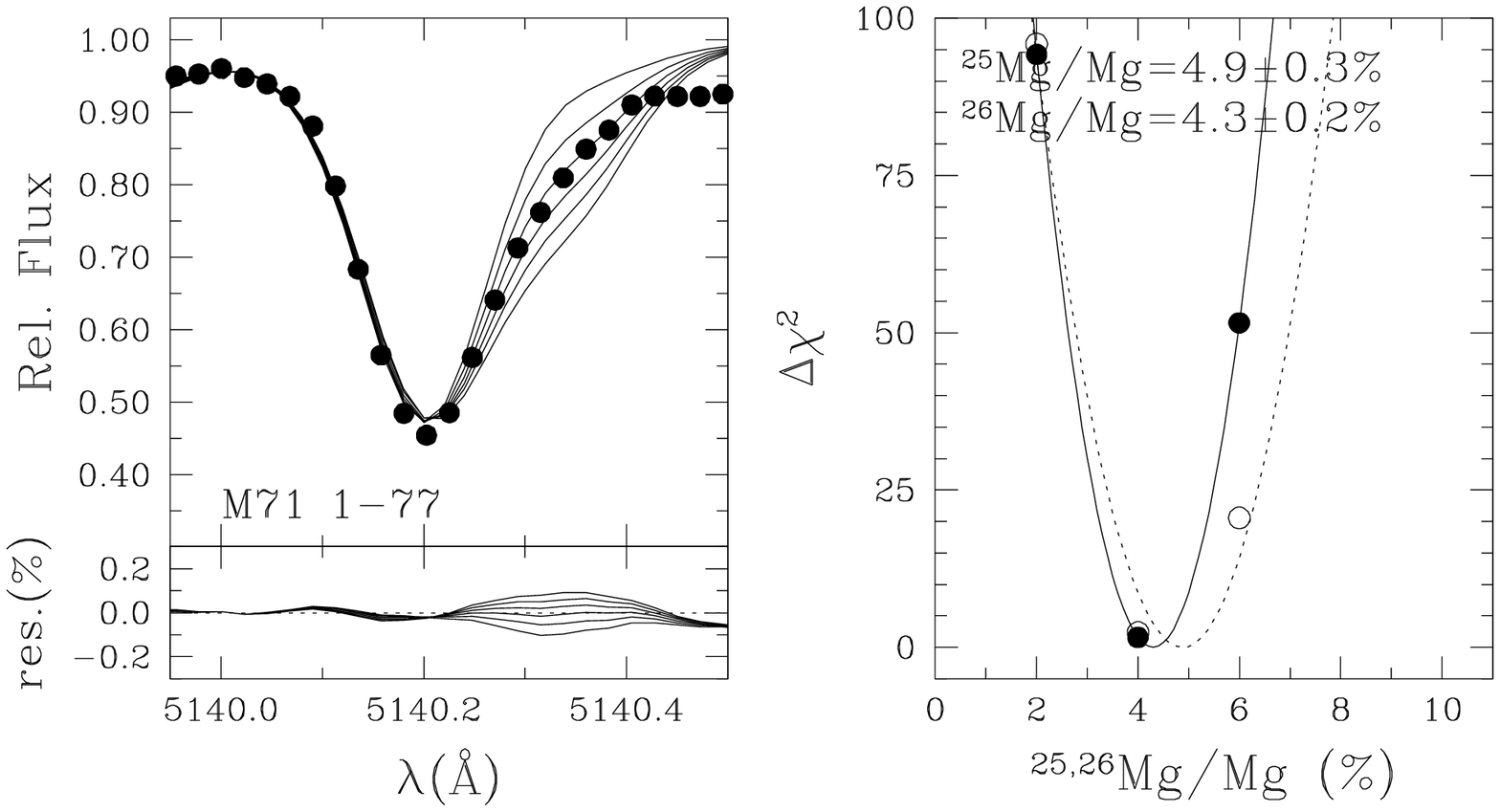}
\caption{Fits for the 5134.3 \AA\ (left), 5138.7 \AA\ (center), and 5140.2 \AA\ (right)
MgH features in the giant M71 1-77. Observed spectra are represented as filled circles,
and synthetic spectra as solid lines. The calculations were performed for 
$^{25,26}$Mg/Mg ratios of 0-10\%. The relative variations of the $\chi^2$ fits are shown
as a function of the isotopic ratio.
}
\label{fits_1-77}
 \end{figure}

\begin{figure}
\includegraphics[width=5.5cm]{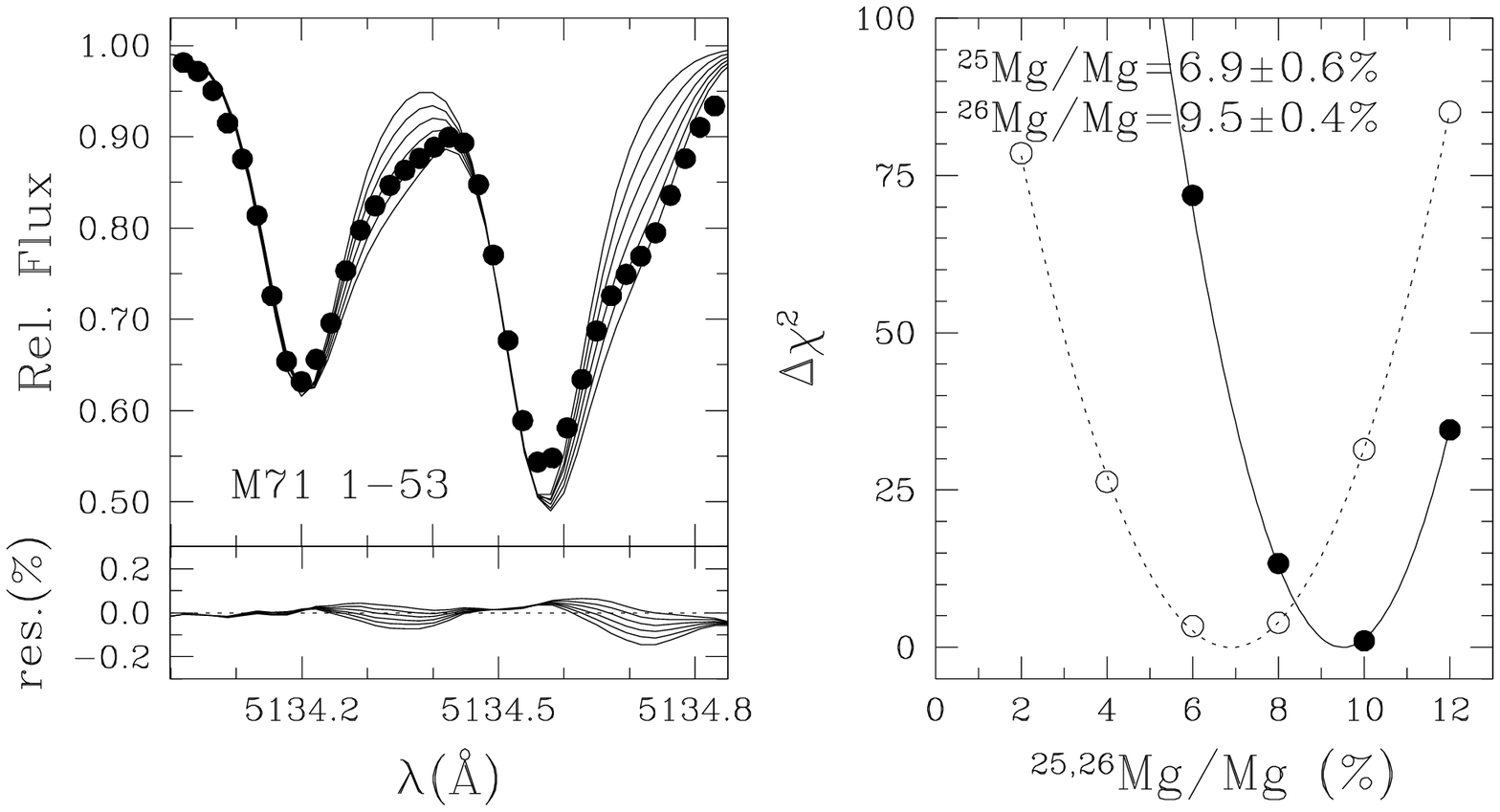}
\includegraphics[width=5.5cm]{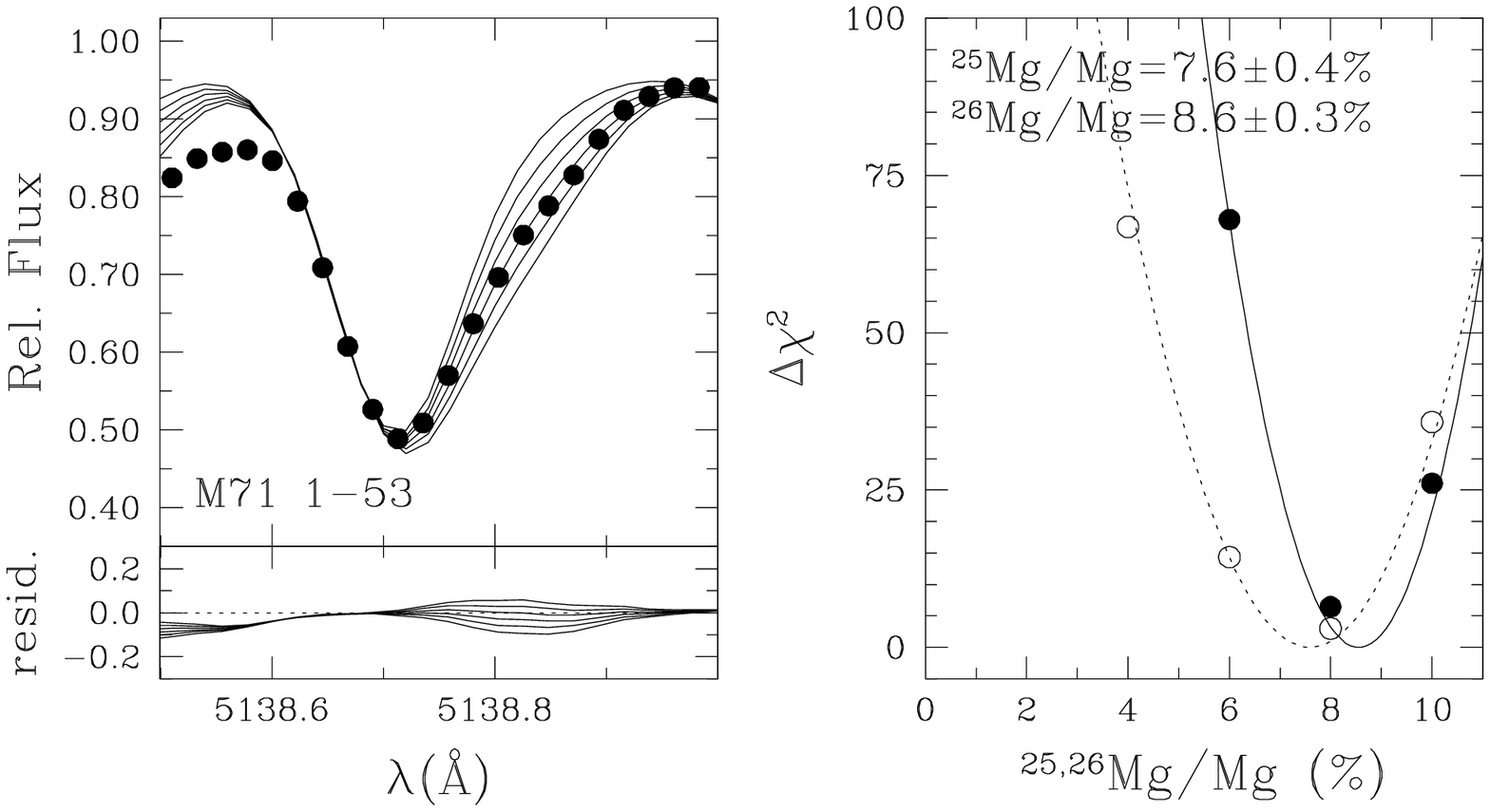}
\includegraphics[width=5.5cm]{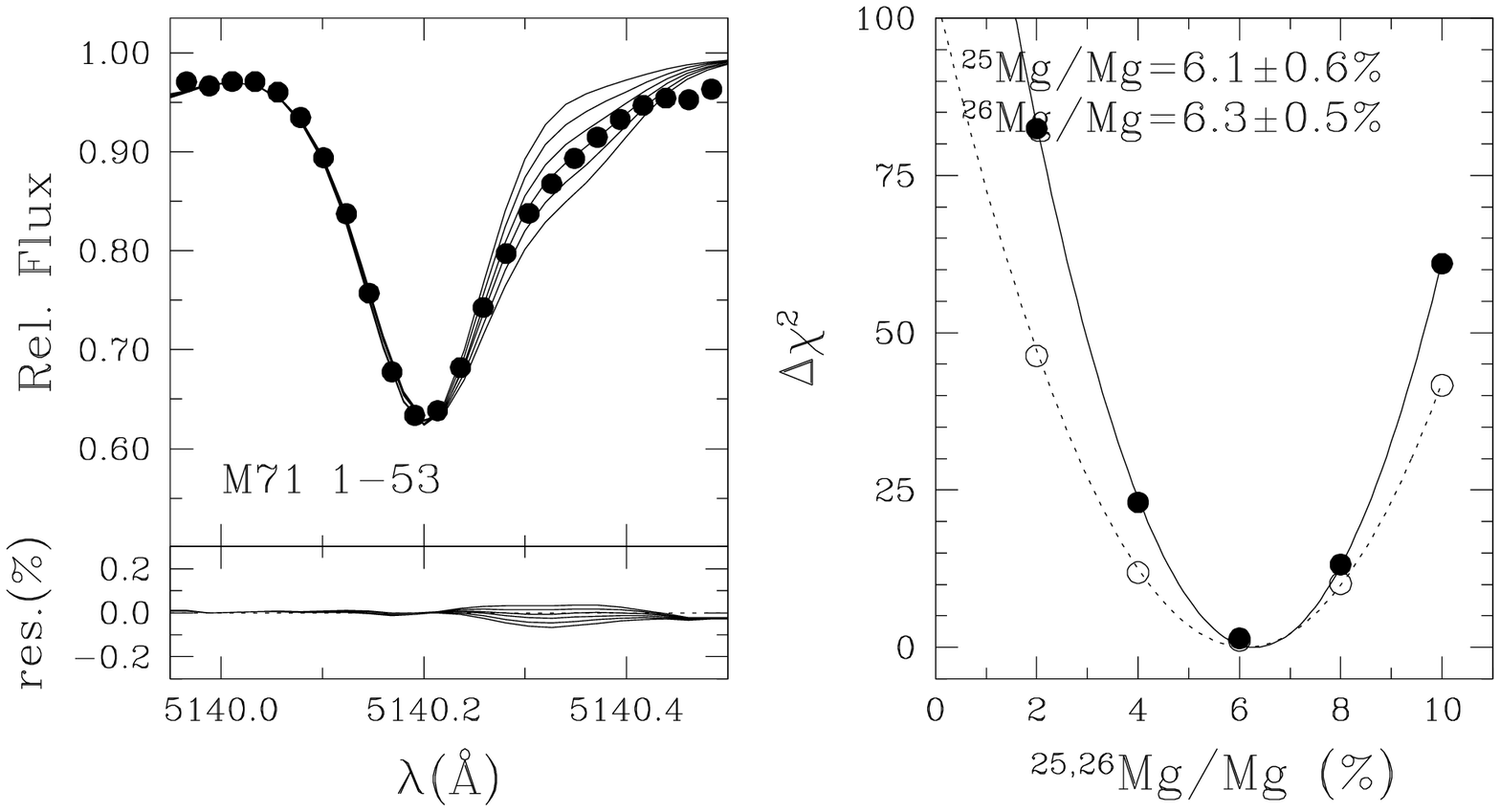}
\caption{Fits for the 5134.3 \AA\ (left), 5138.7 \AA\ (center), and 5140.2 \AA\ (right)
MgH features in the giant M71 1-53. Observed spectra are represented as filled circles,
and synthetic spectra as solid lines. The calculations are shown for 
$^{25,26}$Mg/Mg ratios of 2-12\% (left panel) and 0-10\% (center and right panels). The relative variations of the $\chi^2$ fits are shown
as a function of the isotopic ratio.
}
\label{fits_1-53}
 \end{figure}

\begin{figure}
\includegraphics[width=\hsize]{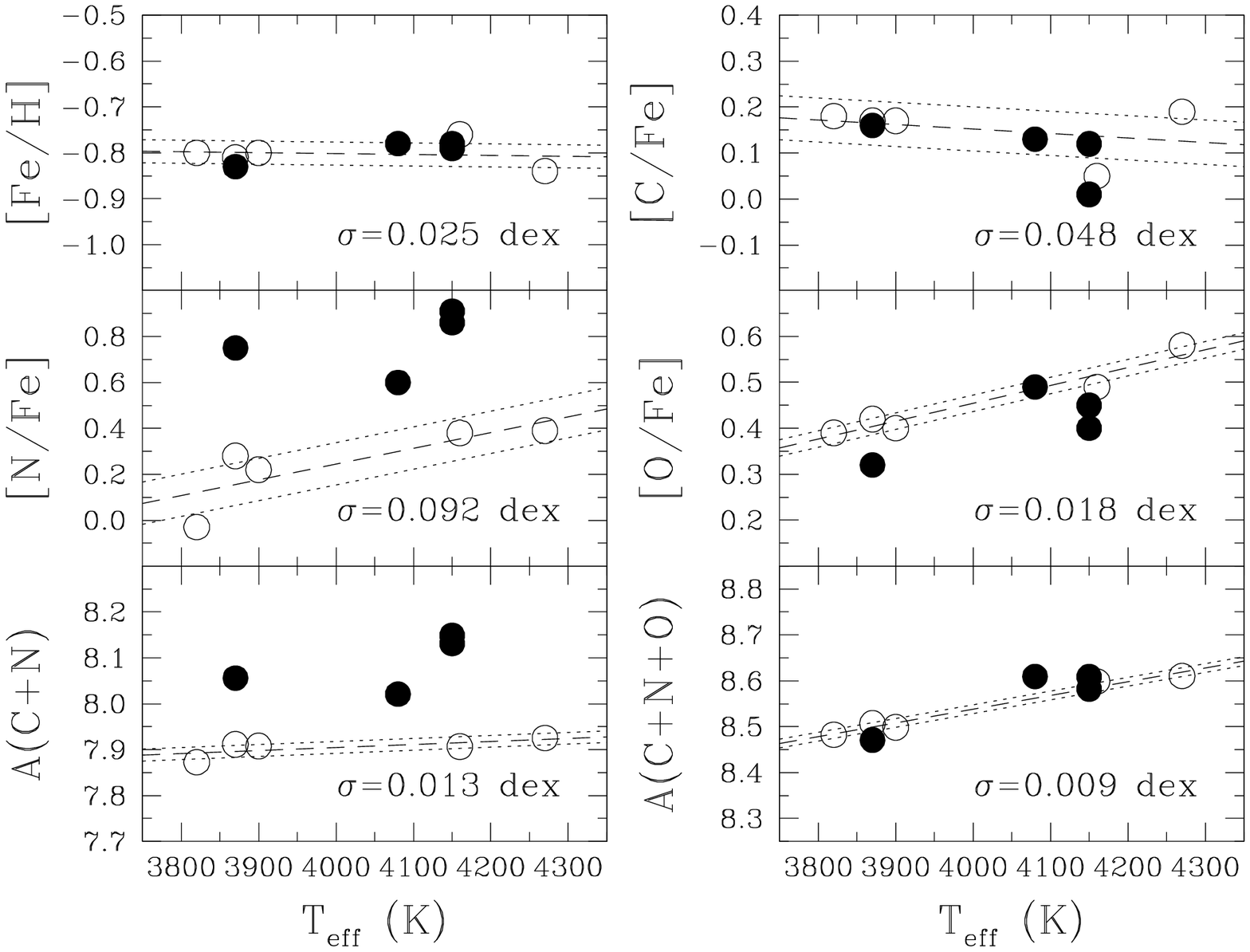}
\caption{[Fe/H], [C,N,O/Fe], A(C+N), and A(C+N+O) abundances (dex) as a function of \tsin.
Open and filled circles represent CN-weak and CN-strong stars, respectively.
The dashed line is a fit to the CN-weak stars, and the dotted lines shows the scatter
($\sigma$) around the fit.
}
\label{oalteff}
 \end{figure}

\begin{figure}
\includegraphics[width=\hsize]{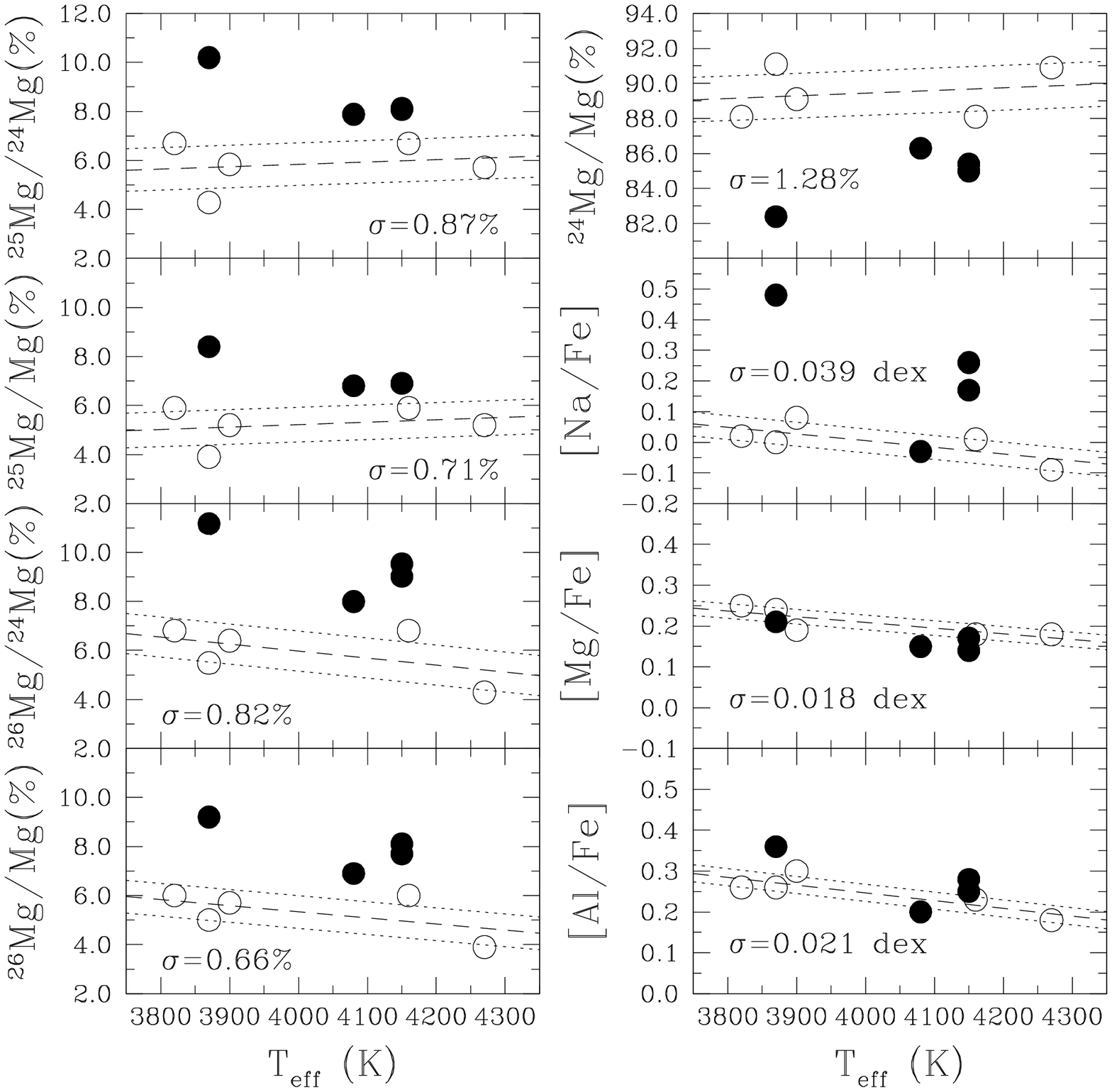}
\caption{$^{24,25,26}$Mg/Mg ratios and [Na,Mg,Al/Fe] abundance ratios (dex) as a function of \tsin.
Open and filled circles represent CN-weak and CN-strong stars, respectively.
The dashed line is a fit to the CN-weak stars, and the dotted lines shows the scatter
($\sigma$) around the fit.
}
\label{oalteff}
 \end{figure}

\begin{figure}
\includegraphics[width=\hsize]{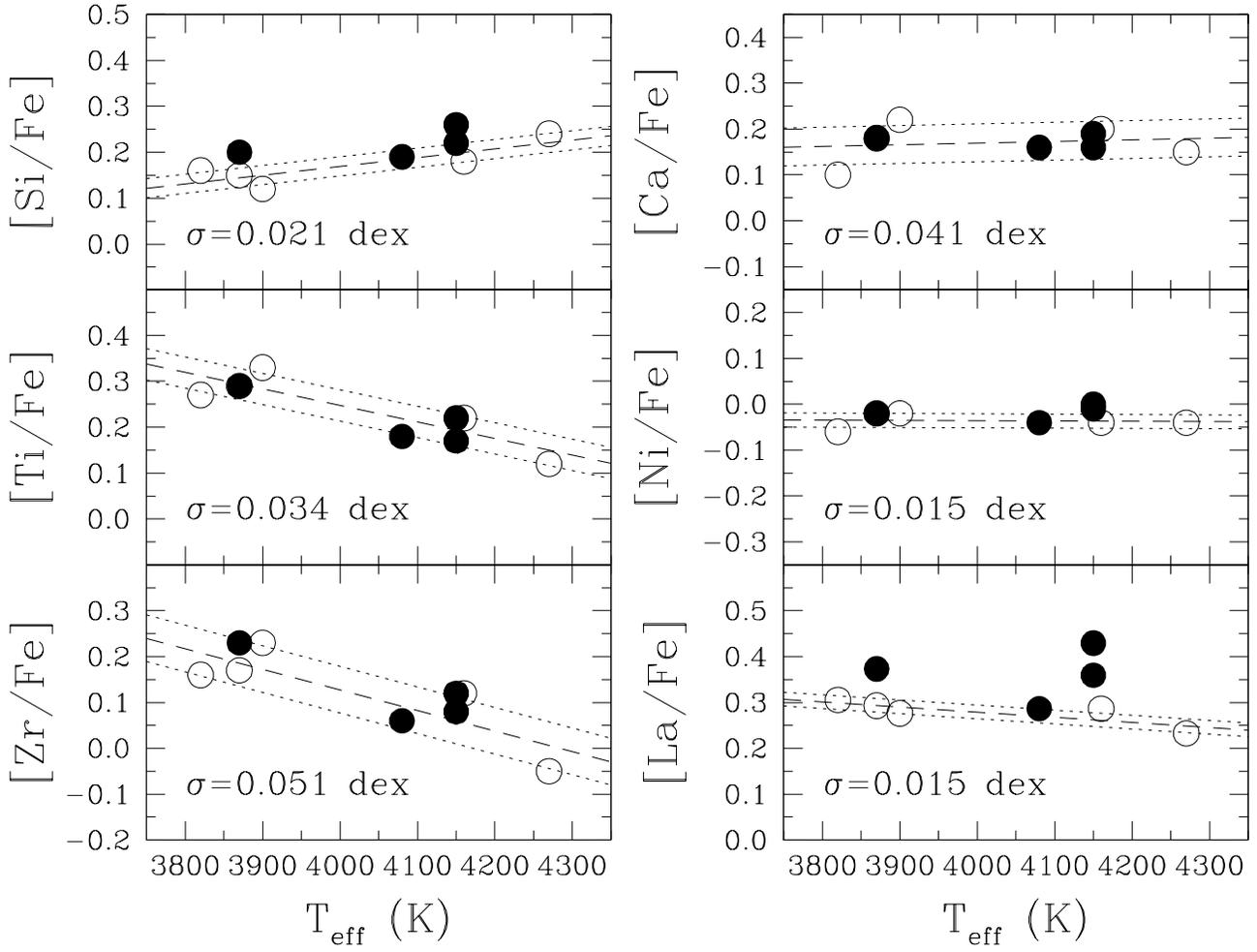}
\caption{[Si,Ca,Ti,Ni,Zr,La/Fe] abundance ratios (dex) as a function of \tsin.
Open and filled circles represent CN-weak and CN-strong stars, respectively.
The dashed line is a fit to the CN-weak stars, and the dotted lines shows the scatter
($\sigma$) around the fit.
}
\label{fezrteff}
 \end{figure}

\begin{figure}
\includegraphics[width=\hsize]{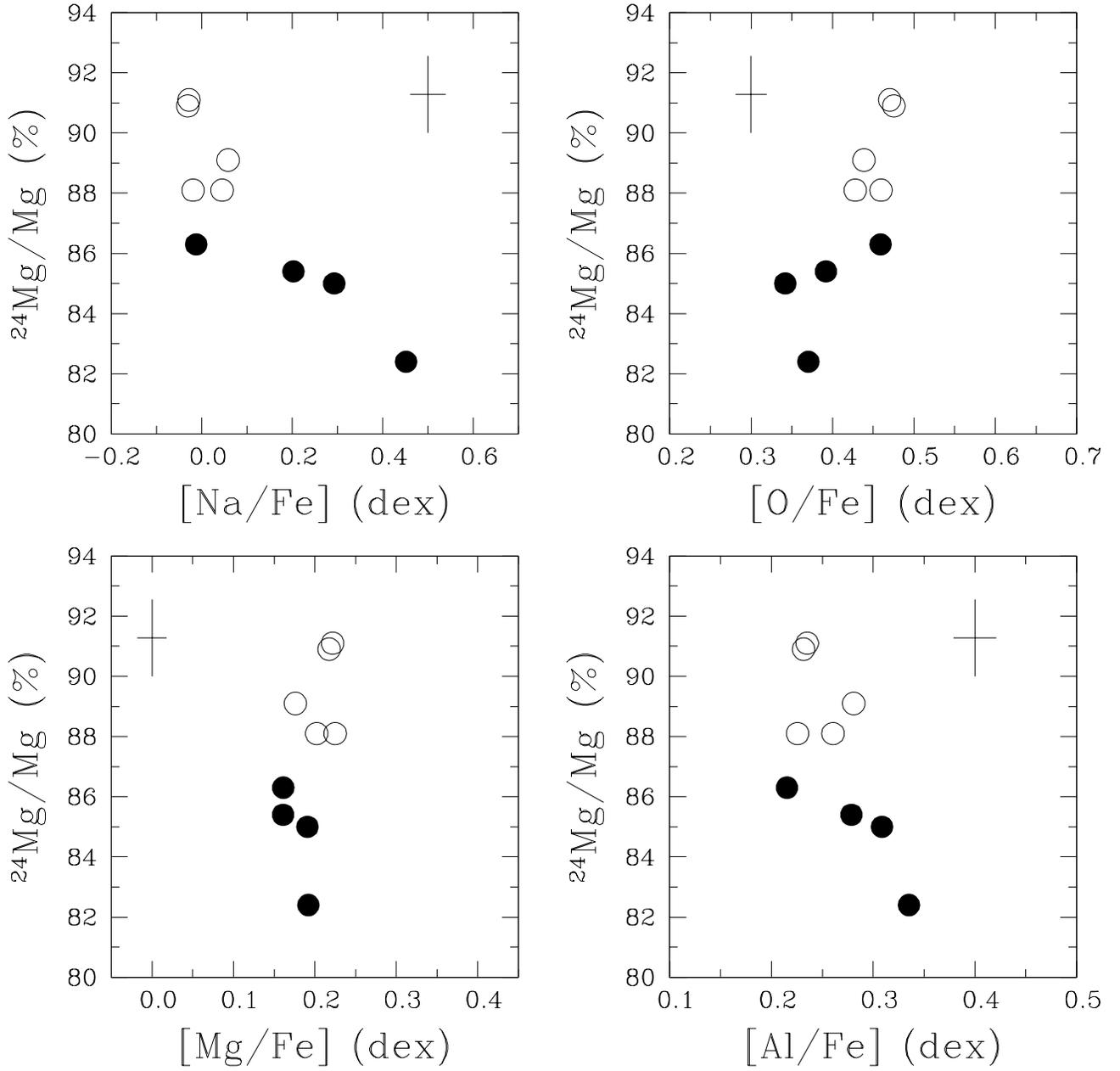}
\caption{$^{24}$Mg/Mg ratios as a function of [Na/Fe], [O/Fe], [Mg/Fe] and [Al/Fe].
Open and filled circles represent CN-weak and CN-strong stars, respectively.
Trends with \teff have been corrected for the elemental [X/Fe] ratios (O, Na, Mg, Al).
}
\label{mg25}
 \end{figure}

\begin{figure}
\includegraphics[width=\hsize]{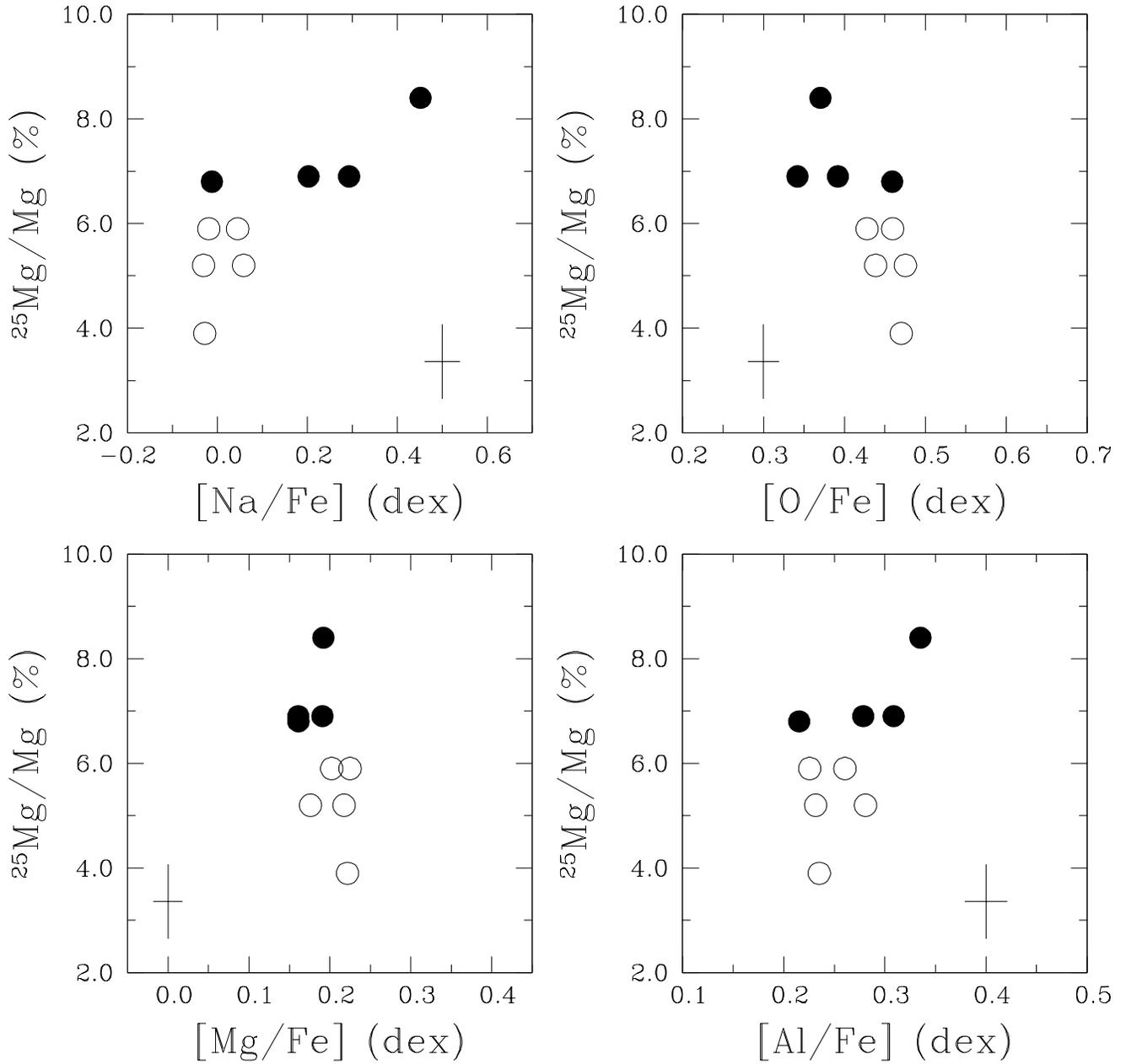}
\caption{$^{25}$Mg/Mg ratios as a function of [Na/Fe], [O/Fe], [Mg/Fe] and [Al/Fe].
Open and filled circles represent CN-weak and CN-strong stars, respectively.
Trends with \teff have been corrected for the elemental [X/Fe] ratios (O, Na, Mg, Al).
}
\label{mg25}
 \end{figure}

\begin{figure}
\includegraphics[width=\hsize]{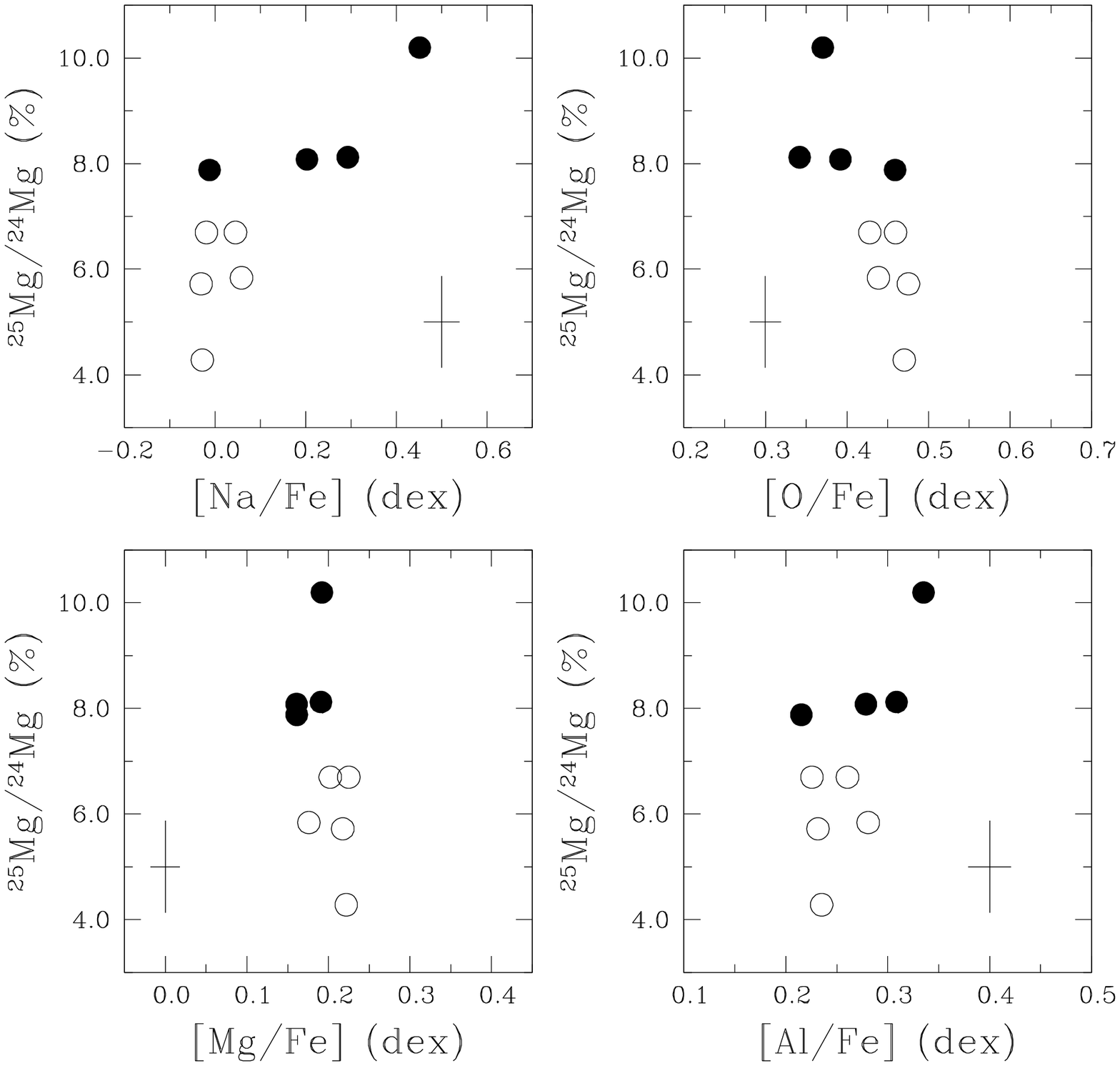}
\caption{$^{25}$Mg/$^{24}$Mg ratios as a function of [Na/Fe], [O/Fe], [Mg/Fe] and [Al/Fe].
Open and filled circles represent CN-weak and CN-strong stars, respectively.
Trends with \teff have been corrected for the elemental [X/Fe] ratios (O, Na, Mg, Al).
}
\label{mg25o24}
 \end{figure}

\begin{figure}
\includegraphics[width=\hsize]{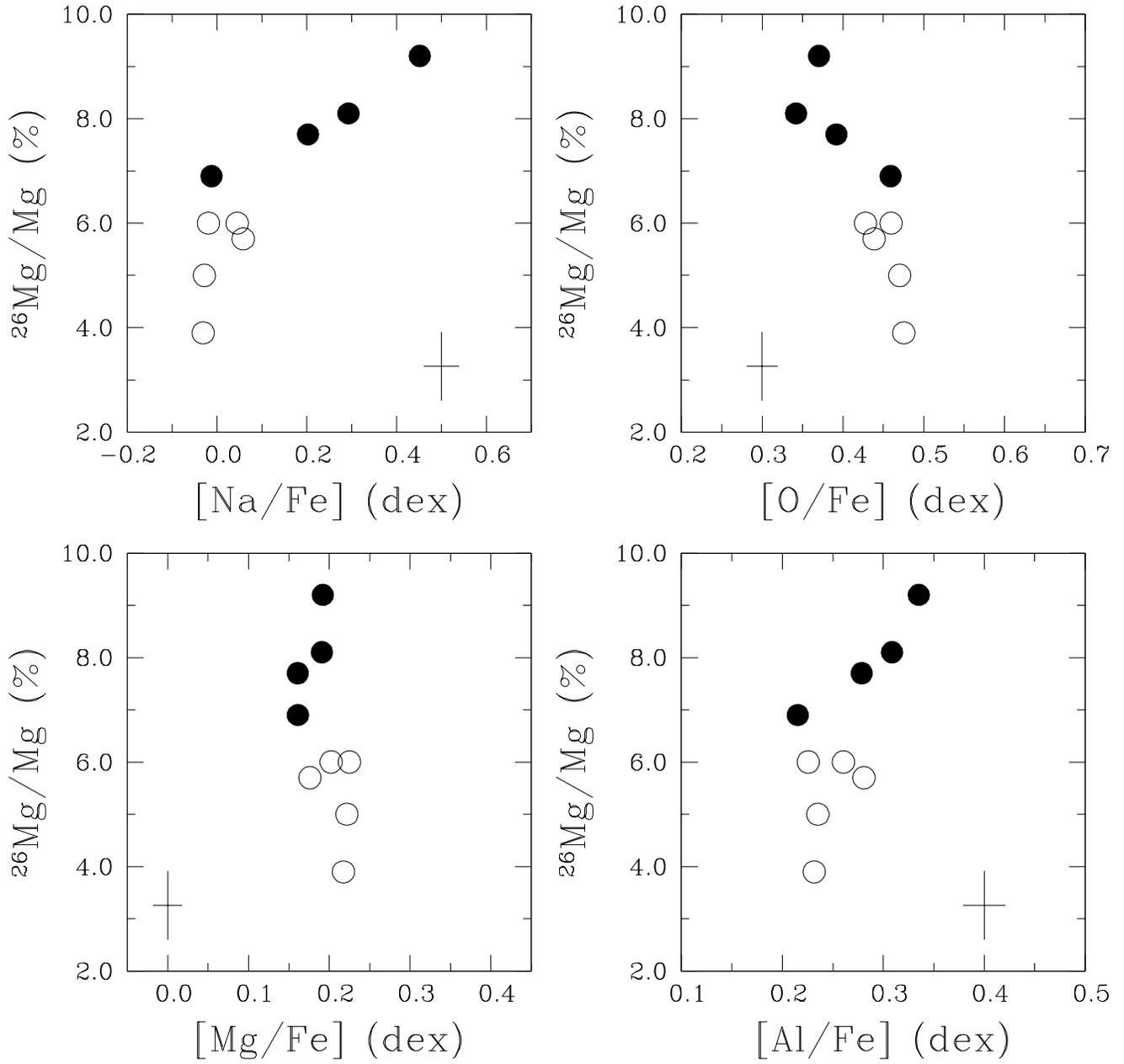}
\caption{$^{26}$Mg/Mg ratios as a function of [Na/Fe], [O/Fe], [Mg/Fe] and [Al/Fe].
Open and filled circles represent CN-weak and CN-strong stars, respectively.
Trends with \teff have been corrected for the elemental [X/Fe] ratios (O, Na, Mg, Al).
}
\label{mg26}
 \end{figure}

\begin{figure}
\includegraphics[width=\hsize]{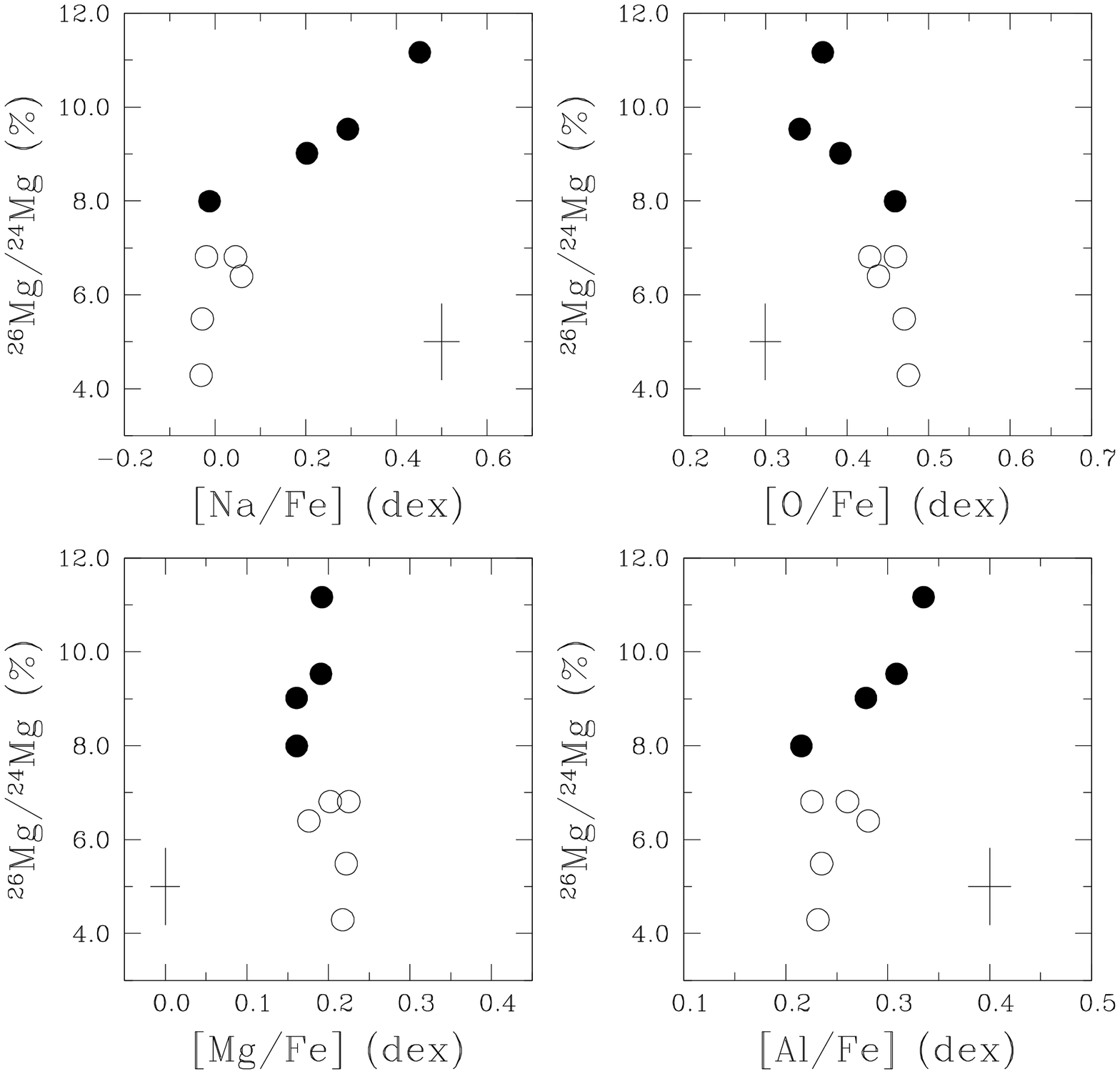}
\caption{$^{26}$Mg/$^{24}$Mg ratios as a function of [Na/Fe], [O/Fe], [Mg/Fe] and [Al/Fe].
Open and filled circles represent CN-weak and CN-strong stars, respectively.
Trends with \teff have been corrected for the elemental [X/Fe] ratios (O, Na, Mg, Al).
}
\label{mg26o24}
 \end{figure}

\begin{figure}
\includegraphics[width=\hsize]{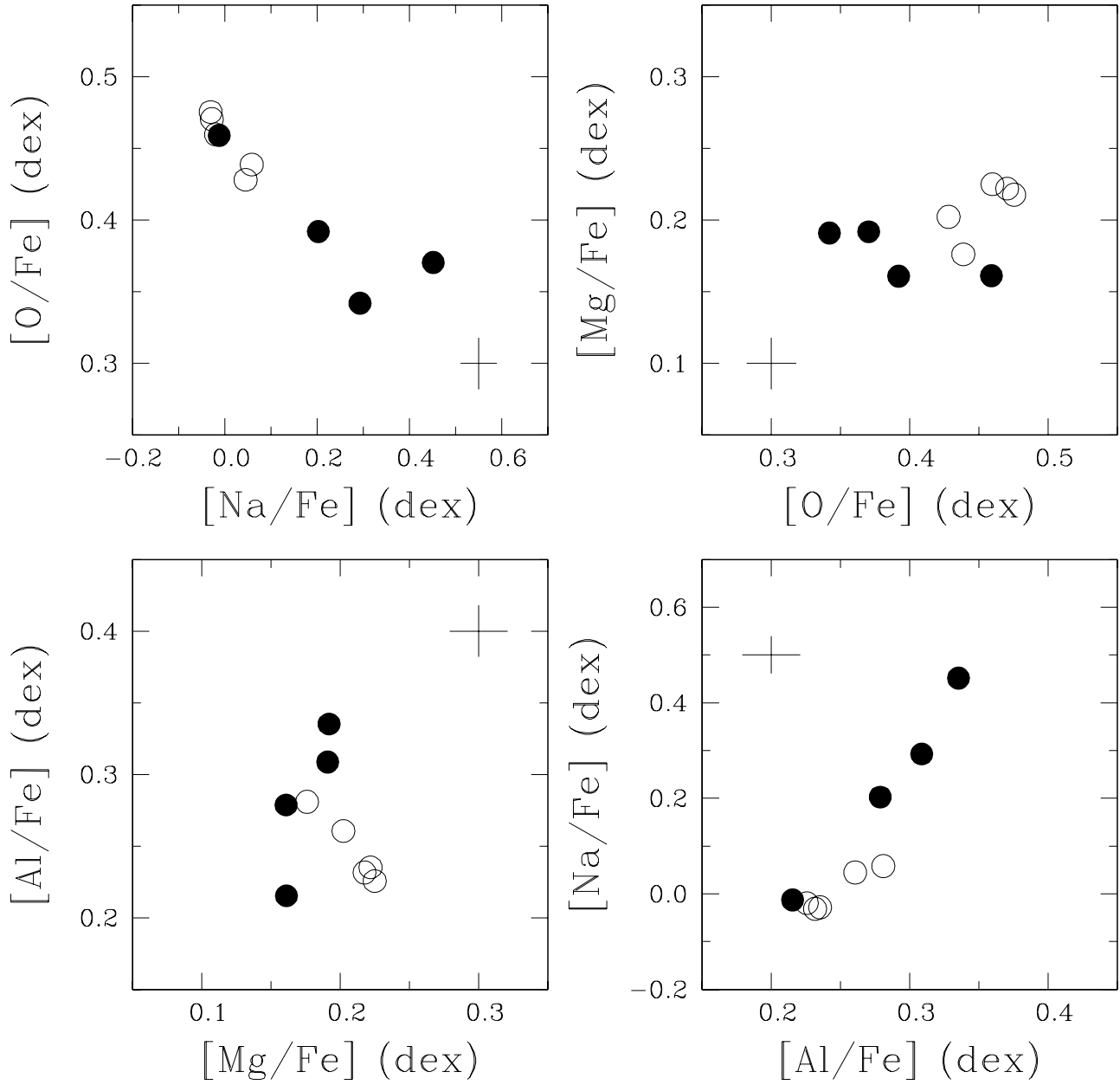}
\caption{Correlations between O, Na, Mg and Al. 
Open and filled circles represent CN-weak and CN-strong stars, respectively.
The abundance ratios have been corrected for trends with \tsin. 
A clear anti-correlation is seen between O and Na, and a
well-defined correlation is seen between Na and Al.
}
\label{onamgal}
 \end{figure}

\begin{figure}
\includegraphics[width=\hsize]{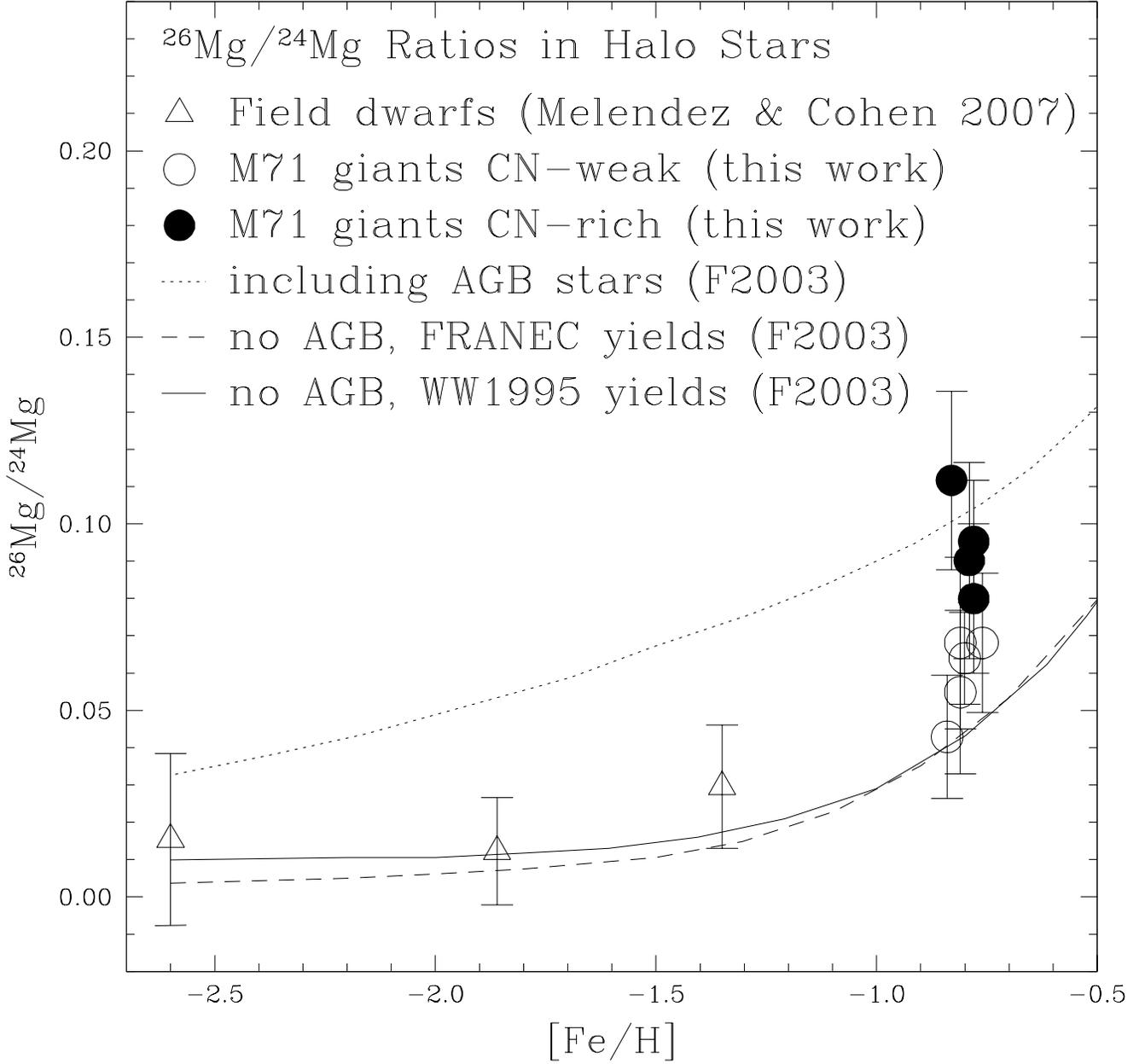}
\caption{Our $^{26}$Mg/$^{24}$Mg ratios in both field
dwarfs (triangles; Melendez \& Cohen 2007) and M71 giants (circles; this work) as
a function of [Fe/H]. 
Chemical evolution models by Fenner et al. (2003) including (dotted line)
and excluding (solid and dashed lines) AGB stars are shown.
At the metallicity of M71 ([Fe/H] = $-0.8$~dex)
the isotopic ratios in the CN-weak stars (open circles) are explained by
massive stars, but the CN-strong stars (filled circles) may have been polluted
by intermediate-mass AGB stars.
}
\label{mg26_feh}
 \end{figure}


\begin{thebibliography}{}
\bibitem[{{Asplund}(2005)}]{2005ARA&A..43..481A}
{Asplund}, M. 2005, \araa, 43, 481

\bibitem[Briley 
\& Cohen(2001)]{2001AJ....122..242B} Briley, M.~M., \& Cohen, J.~G.\ 2001, \aj, 122, 242 

\bibitem[Busso et al(2001)]{busso01}
Busso, M., Gallino, R., Lambert, D.~L., Travaglio, C. \& Smith, V.~V.,
2001,\apj, 557, 802

\bibitem[Carretta et al(2004)]{carretta04}
Carretta, E.,  Gratton, R.~G., Bragaglia, A., Bonifacio, P. \& Pasquini, L.,
2004, \aap, 416, 925


\bibitem[Carretta et al(2007)]{carretta07}
Carretta, E., Recio-Blanco, A., Gratton, R.~G.,  Piotto, G.
\& Bragaglia, A., 2007, \apjl, 671, L125 

\bibitem[Castelli et al.(1997)]{1997A&A...318..841C} 
Castelli, F., Gratton, R.~G., \& Kurucz, R.~L.\ 1997, \aap, 318, 841 

\bibitem[Cohen, Behr \& Briley(2001)]{cohen01}
Cohen, J.~G., Behr, B.~B. \& Briley, M.~M., 2001, \aj, 122, 1420

\bibitem[Cohen 
\& Mel{\'e}ndez(2005)]{2005AJ....129..303C} Cohen, J.~G., \& Mel{\'e}ndez, J.\ 2005, \aj, 129, 303 


\bibitem[Cohen, Briley \& Stetson(2005)]{cohen_m15_c}
Cohen, J.~G., Briley, M.~M. \& Stetson, P.~B., 2005, \aj, 130, 1177

\bibitem[Collet et al.(2007)]{2007A&A...469..687C} 
Collet, R., Asplund, M., \& Trampedach, R.\ 2007, \aap, 469, 687 

\bibitem[Cutri et al(2003)]{2mass2} 
Cutri, R.~M. et al, 2003,
``Explanatory Supplement to the 2MASS All-Sky Data Release,
http:\\www.ipac.caltech.edu/2mass/releases/allsky/doc/explsup.html

\bibitem[{{da Silva} {et~al.}(2006){da Silva}, {Girardi}, {Pasquini},
  {Setiawan}, {von der L{\"u}he}, {de Medeiros}, {Hatzes}, {D{\"o}llinger}, \&
  {Weiss}}]{2006A&A...458..609D}
{da Silva}, L., {Girardi}, L., {Pasquini}, L., {et~al.} 2006, \aap, 458, 609

\bibitem[Decressin et al(2007)]{decressin07}
Decressin, T., Meynet, G., Charbonnel, C., Prantzos, N. \& Ekstrom, S.,
2007, \aap, 464, 1029

\bibitem[{{Demarque} {et~al.}(2004){Demarque}, {Woo}, {Kim}, \&
  {Yi}}]{2004ApJS..155..667D}
{Demarque}, P., {Woo}, J.-H., {Kim}, Y.-C., \& {Yi}, S.~K. 2004, \apjs, 155,
  667
  
\bibitem[Denisenkov \& Denisenkova(1990)]{dennisenkov}
Denisenkov, P.~A. \& Denisenkova, S.~N., 1990, Soviet Astronomy Letters, 16, 275  

\bibitem[Fenner et al(2003)]{fenner03}
Fenner, Y. et al, 2003, PASA, 20, 340

\bibitem[Gay \& Lambert(2000)]{2000ApJ...533..260G} 
Gay, P.~L., \& Lambert, D.~L.\ 2000, \apj, 533, 260

\bibitem[Gratton et al(2001)]{gratton01}
Gratton, R. et al, 2001, \aap, 369, 87


\bibitem[Gratton, Sneden \& Carretta(2004)]{gratton_araa}
Gratton, R., Sneden, C. \& Carretta, E., 2004, \araa, 42, 385

\bibitem[Harris(1996)]{harris96} 
Harris, W.~E.\ 1996, \aj, 112, 1487


\bibitem[Hekker \& Mel{\'e}ndez(2007)]{2007A&A...475.1003H} 
Hekker, S., \& Mel{\'e}ndez, J.\ 2007, \aap, 475, 1003 

\bibitem[Hirschi et al.(2008)]{hirschi08} Hirschi, R., et al.\ 
2008, 10th Symposium on Nuclei in the Cosmos, in press, arXiv:0811.4654 


\bibitem[Karakas \& Lattanzio(2003)]{karakas03}
Karakas, A.~I. \& Lattanzio, J.~C., 2003, PASA, 20, 279

\bibitem[Karakas 
\& Lattanzio(2007)]{karakas08} Karakas, A., \& Lattanzio, J.~C.\ 2007, PASA, 24, 103 


\bibitem[Kroupa \& Boily(2002)]{kroupa02}
Kroupa,~P. \& Boily, C.~M., 2002, \mnras, 336, 1188

\bibitem[Lawler et al.(2001)]{lawler01} Lawler, J.~E., 
Bonvallet, G., \& Sneden, C.\ 2001, \apj, 556, 452 

\bibitem[Lee(2005)]{2005JKAS...38...23L} Lee, S.-G.\ 2005, Journal of 
Korean Astronomical Society, 38, 23 

\bibitem[McWilliam 
\& Lambert(1988)]{1988MNRAS.230..573M} McWilliam, A., \& Lambert, D.~L.\ 1988, \mnras, 230, 573 

\bibitem[Mel{\'e}ndez \& Barbuy(1999)]{1999ApJS..124..527M} 
Mel{\'e}ndez, J., \& Barbuy, B.\ 1999, \apjs, 124, 527 

\bibitem[Mel{\'e}ndez \& Cohen(2007)]{2007ApJ...659L..25M} 
Mel{\'e}ndez, J., \& Cohen, J.~G.\ 2007, \apjl, 659, L25 

\bibitem[Mel{\'e}ndez et al.(2008)]{2008A&A...484L..21M} 
Mel{\'e}ndez, J., et al.\ 2008, \aap, 484, L21 

\bibitem[Mel{\'e}ndez \& Asplund(2008)]{melendez_asplund}
Mel{\'e}ndez, J. \& Asplund, M., 2008, \aap, 490, 817

\bibitem[Melendez \& Barbuy(2009)]{melendezbarbuy09} Melendez, J. \& Barbuy, B.\ 2009, \aap, in press, arXiv:0901.4451

\bibitem[Plez 
\& Cohen(2005)]{2005A&A...434.1117P} Plez, B., \& Cohen, J.~G.\ 2005, \aap, 434, 1117 

\bibitem[Plez et al.(2008)]{plez2008} Plez, B. et al.\ 2008, 
14th Cambridge Workshop on Cool Stars, Stellar Systems, and the Sun, 384, poster

\bibitem[Ram{\'{\i}rez} \& Cohen(2002)]{ramirez02}
Ram{\'{\i}}rez, S. \& Cohen, J.~G., 2002, \aj, 123, 3277

\bibitem[{{Ram{\'{\i}}rez} \& {Mel{\'e}ndez}(2005)}]{2005ApJ...626..465R}
{Ram{\'{\i}}rez}, I. \& {Mel{\'e}ndez}, J. 2005, \apj, 626, 465

\bibitem[Shetrone(1996)]{shetrone96}
Shetrone, M.~D., 1996, \aj, 112, 2639

\bibitem[Skrutskie et al(1997)]{2mass1}
Skrutskie, M.~F., Schneider, S.E., Stiening, R., Strom, S.E.,
Weinberg, M.D., Beichman, C., Chester, T. et al, 1997, in {\it{The
Impact of Large Scale Near-IR Sky Surveys}}, ed. 
F.Garzon et al (Dordrecht: Kluwer), p. 187

\bibitem[Smith \& Norris(1982)]{1982ApJ...254..149S} 
Smith, G.~H., \& Norris, J.\ 1982, \apj, 254, 149 

\bibitem[Smith et al.(2005)]{2005ApJ...633..392S} Smith, V.~V., Cunha, K., 
Ivans, I.~I., Lattanzio, J.~C., Campbell, S., 
\& Hinkle, K.~H.\ 2005, \apj, 633, 392 


\bibitem[{{Sneden}(1973)}]{1973PhDT.......180S}
{Sneden}, C.~A. 1973, PhD thesis, AA(THE UNIVERSITY OF TEXAS AT AUSTIN.)

\bibitem[Stetson(2000)]{2000PASP..112..925S} 
Stetson, P.~B.\ 2000,  \pasp, 112, 925

\bibitem[Travaglio et al(2004)]{travaglio04}
Travaglio, C., Gallino, R., Arnone, E., Cowan, J.~C., Jordan, F.
\& Sneden, C., 2004, \apj, 601, 864

\bibitem[Ventura \& D'Antona(2008)]{ventura08}
Ventura, P. \& D'Antona, F.~D., 2008, \mnras, 385, 2034

\bibitem[Vogt et al.(1994)]{1994SPIE.2198..362V} Vogt, S.~S., et al.\ 1994, 
\procspie, 2198, 362 

\bibitem[Woosley \& Weaver(1995)]{woosley95}
Woosley, S.~E. \& Weaver, T.~A., 1995, \apj, 101, 181

\bibitem[Yong et al.(2003)]{2003A&A...402..985Y} 
Yong, D., Grundahl, F., Lambert, D.~L., Nissen, P.~E., \& Shetrone, M.~D.\ 2003, \aap, 402, 985 

\bibitem[Yong et al.(2006)]{2006ApJ...638.1018Y} 
Yong, D., Aoki, W., \& Lambert, D.~L.\ 2006, \apj, 638, 1018 

\bibitem[Yong et al.(2008)]{2008ApJ...689.1020Y} Yong, D., Mel{\'e}ndez, 
J., Cunha, K., Karakas, A.~I., Norris, J.~E., 
\& Smith, V.~V.\ 2008, \apj, 689, 1020 

\bibitem[Yong et al.(2009)]{2009arXiv0902.1773Y} Yong, D., Grundahl, F., 
D'Antona, F., Karakas, A.~I., Lattanzio, J.~C., 
\& Norris, J.~E.\ 2009, ApJ Letters, in press, arXiv:0902.1773 

\end{thebibliography}
\end{document}